\documentclass[twocolumn]{aastex62}
\usepackage{amsmath}

\graphicspath{{./}{figures/}}

\received{January 1, 2020}
\accepted{July 18, 2020}

\shorttitle{Detecting Exoplanets with Eclipsing Binaries}
\shortauthors{Bellotti et al.}

\begin{document}

\title{Detecting Exoplanets Using Eclipsing Binaries as Natural Starshades}

\correspondingauthor{Stefano Bellotti}
\email{info.stefanobellotti@gmail.com}

\author[0000-0002-2558-6920]{Stefano Bellotti}
\affil{DARK, Niels B\"ohr Institute, University of Copenhagen, Lyngbyvej 2, 4. sal, 2100 Copenhagen \O, Denmark}
\affil{Universit\'e de Toulouse, UPS-OMP, IRAP, 14 avenue E. Belin, Toulouse, F-31400 France}
\affil{CNRS, IRAP/UMR 5277, Toulouse, 14 avenue E. Belin, F-31400 France}

\author[0000-0001-6047-8469]{Ann I. Zabludoff}
\affil{Steward Observatory, University of Arizona, 933 North Cherry Avenue, Tucson AZ 85721}

\author[0000-0002-4951-8025]{Ruslan Belikov}
\affil{NASA Ames Research Center, Moffett Field, CA 94035, USA}

\author[0000-0002-1097-9908]{Olivier Guyon}
\affil{Steward Observatory, University of Arizona, 933 North Cherry Avenue, Tucson AZ 85721}
\affil{NAOJ}

\author{Chirag Rathi}
\affil{Steward Observatory, University of Arizona, 933 North Cherry Avenue, Tucson AZ 85721}



\begin{abstract}
We investigate directly imaging exoplanets around eclipsing binaries, using the eclipse as a natural tool for dimming the binary and thus increasing the planet to star brightness contrast. At eclipse, the binary becomes point-like, making coronagraphy possible. We select binaries where the planet-star contrast would be boosted by $>10\times$ during eclipse, making it possible to detect a planet that is $\gtrsim10\times$ fainter or in a star system that is $\sim2$-$3\times$ more massive than otherwise. Our approach will yield insights into planet occurrence rates around binaries versus individual stars. We consider both  self-luminous (SL) and reflected light (RL) planets. In the SL case, we select binaries whose age is young enough so that an orbiting SL planet would remain luminous; in U Cep and AC Sct, respectively, our method is sensitive to SL planets of $\sim$4.5$M_J$ and $\sim$9$M_J$ with current ground- or near-future space-based instruments, and $\sim$1.5$M_J$ and $\sim$6$M_J$ with future ground-based observatories. In the RL case, there are three nearby ($\lesssim50$ pc) systems---V1412 Aql, RR Cae, RT Pic---around which a Jupiter-like planet at a planet-star separation of $\gtrsim20$ mas might be imaged with future ground- and space-based coronagraphs. A Venus-like planet at the same distance might be detectable around RR Cae and RT Pic. A habitable Earth-like planet represents a challenge; while the planet-star contrast at eclipse and planet flux are accessible with a 6-8m space telescope, the planet-star separation is 1/3 - 1/4 of the angular separation limit of modern coronagraphy.

\end{abstract}


\section{Introduction}\label{sec:intro}

Coronagraphs, nulling interferometry, and man-made starshades are the existing strategies for imaging exoplanets directly. Is there a way to dramatically improve these techniques? Here we consider using the eclipse in an eclipsing binary system to dim the observed brightness of the primary and increase the planet to star flux contrast, i.e., we explore the possibility of employing a {\it natural} starshade as a tool to find additional exoplanets around binaries via direct imaging, along with the mentioned techniques.

The NASA Exoplanet Archive \citep{2013PASP..125..989A} lists 270 binary systems with exoplanets, but only six binaries have planets detected via direct imaging \citep{2010ApJ...725.1405B,2011AJ....141..119K,2014ApJ...787..104C,2014ApJ...781...20K,2015ApJ...804...96G,2019A&A...626A..99J}. All six are self-luminous planets with minimum masses intermediate between 6 and 20 $M_J$. Our alternate direct imaging method can explore a new parameter space by 1) targeting binaries or stars that are unusual compared to previously observed exoplanet systems and 2) making different (fainter) types of planets accessible.

Planets in binary systems could represent an important fraction of planet demography, especially given that $\sim 45\%$ of Sun-like stars in the Galactic field are part of a multiple system \citep{2010ApJS..190....1R}. \citet{2013MNRAS.436..650P} estimate statistically the percentage of Solar System analogues, defined either as an individual G dwarf or a binary system with separation $>100$-300 AU and a G dwarf component, that host exoplanets. This percentage declines from 65-95\% to 20-65\% from 1 to 100 AU for planets on circumprimary (S-type) orbits, whereas it increases from 5-59\% to 34-75\% from 1 to 100 AU for planets on circumbinary (P-type) orbits.

The effect of binarity, relative to single stars, on planet occurrence rates is uncertain. On one hand, the proximity of a stellar companion could induce disk truncation \citep{2015ApJ...799..147J} and suppress planet formation \citep{2019arXiv191201699M}, reducing (by $0.3\times$) the occurrence rate compared to that in wider binary or individual star systems \citep{2016AJ....152....8K}. On the other hand, \citet{2018AJ....156...31M} do not observe this suppression within $\simeq50$ AU, and some planets have been discovered orbiting in S-type configurations within tight binary systems \citep[e.g.,][]{2015pes..book..309T}. Furthermore, the material-rich environments that form massive stars and binaries may readily produce high-mass protoplanetary disks and then gas giant planets \citep{2008ApJ...673..502K}.

\citet{2013A&A...549A..95Z} analyze detached post-common-envelope binaries and find that 90\% of those observed for $\sim$5 years have eclipse timing variations that could be explained by a circumbinary companion. The Search for Planets Orbiting Two Stars (SPOTS) survey  \citep{2014A&A...572A..91T,2016A&A...593A..38B,2018A&A...619A..43A} constrains the frequency of wide ($<$ 1000 AU) orbit substellar companions to between 0.9\% and 9\%, consistent with that around single stars. The combination of the low rate from the SPOTS survey and the high frequency from the \citet{2013A&A...549A..95Z} study suggests a second generation scenario of planet formation around post-common-envelope binaries, i.e., the planet forms after the binary. In this context, our approach has the potential not only to yield further insights on planet occurrence rates around binaries, but also to differentiate among theories of planet formation in binary environments.

Using eclipsing binaries to image exoplanets directly could also expand our knowledge of the kinds of binaries around which planets can form and evolve. For example, while planets have been discovered around eclipsing binaries using the eclipse timing method (HW Vir, \citealt{2009AJ....137.3181L}; DP Leo, \citealt{2010ApJ...708L..66Q}; NN Ser, \citealt{2009ApJ...706L..96Q,2010A&A...521L..60B}; NY Vir, \citealt{2012ApJ...745L..23Q}; RR Cae,
\citealt{2012MNRAS.422L..24Q}), the host properties are narrow and biased. The hosts are generally short-period compact binaries with a low-mass star or a white dwarf component, as the eclipse minimum can be timed more precisely in these cases. Instead, for our method to work efficiently, we require that the dip in magnitude at eclipse is large enough to yield a substantial gain in contrast, regardless of the type of components.

\begin{figure*}[t]
	\includegraphics[width=\textwidth]{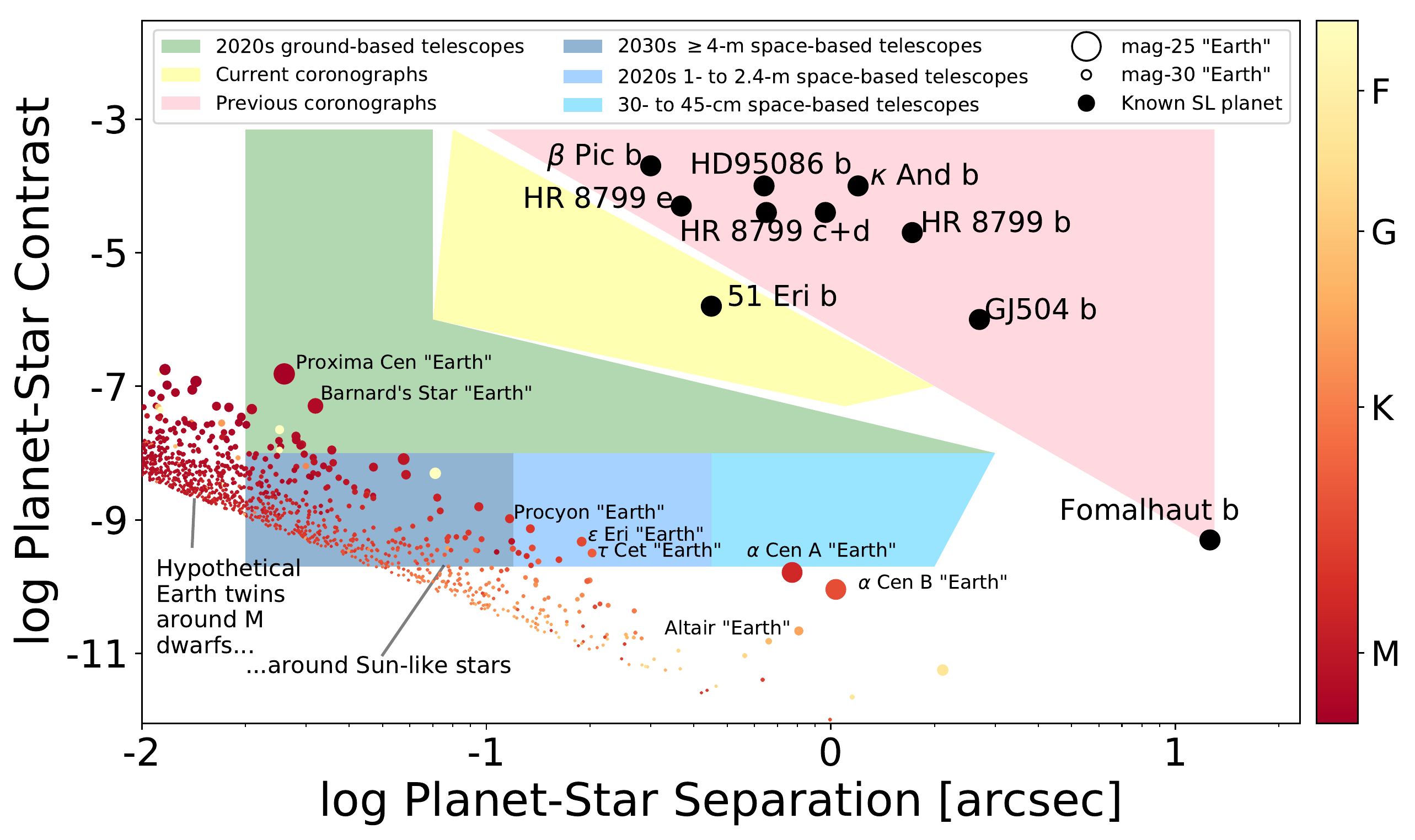}
	\caption{Planet-to-star contrast versus separation plane for the direct imaging method. The capabilities of existing or planned instruments are bundled into different technology zones (colored regions): in pink, VLT-NACO \citep{2003SPIE.4841..944L,2003SPIE.4839..140R}, Subaru-HiCIAO \citep{2006SPIE.6269E..0VT}, Keck-NIRC2 \citep{2000SPIE.4008....2M}; in yellow, VLT-SPHERE \citep{2008SPIE.7014E..18B}, Gemini-GPI \citep{2014SPIE.9148E..0JM}, \textit{JWST}-NIRCam \citep{2007SPIE.6693E..0HK}; in green, EELT-EPIC \citep{2008SPIE.7015E..1SK}, TMT-PSI \citep{2018SPIE10703E..0ZG}, TMT-PFI and SEIT \citep{2013A&A...551A..99C}; in blue, \textit{WFIRST}-CGI \citep{2018SPIE10698E..6PB}, \textit{LUVOIR}-ECLIPS \citep{2019AAS...23314806J} and \textit{HabEx} \citep{2020arXiv200106683G}. The gap between 30- to 45-cm space based telescopes (lightest blue) and previous coronagraphs (pink) is not significant. The groups of instruments defining each colored technology zone vary in their wavelength coverage, so our comparisons to them are rough guides to what is possible. Here, we plot self-luminous planets already discovered through direct imaging with infrared coronagraphs (black points) as well as simulated ``Earths'' that could be directly imaged in reflected light (colored points). The ``Earths'' are assumed to lie around a sample of M--F type stars within 20 pc; their color encodes the spectral type of the host, while their size is scaled according to the apparent infrared magnitude of the planet. Not surprisingly, no ``Earths'' falls within the reach of existing instruments (yellow).}
	\label{fig:Belikov}
\end{figure*}

Here we present the application of the natural-starshade method to both self-luminous (SL) and reflected light (RL) planets, in order to assess whether eclipsing binaries represent competitive targets for potential detections with current or future imaging technology. Figure~\ref{fig:Belikov} illustrates the observational space and depicts previous, current, and future imaging facilities according to their actual or expected performances. 
Some directly imaged self-luminous exoplanets are shown in Figure~\ref{fig:Belikov} for comparison: $\beta$ Pic b \citep{2009A&A...493L..21L,2010Sci...329...57L}, HD 95086 b \citep{2013ApJ...779L..26R}, $\kappa$ And b \citep{2013AAS...22132406C}, HR 8799 b,c,d,e \citep{2008Sci...322.1348M,2010Natur.468.1080M}, 51 Eri b \citep{2015Sci...350...64M}, GJ 504 b \citep{2013ApJ...774...11K}, and Fomalhaut b \citep{2008Sci...322.1345K}, which could be surrounded by a cloud of dust or a disk \citep{2013ApJ...769...42G,2015ESS.....320107L}, or due to a massive collision of two planetesimals \citep{2020arXiv200408736G}.

We also simulate the presence of an Earth-twin (i.e., same radius and albedo as Earth, and flux received from the binary equal to the solar constant) around a sample of nearby (within 20 pc) stars ranging from M to F type. No datapoint falls in the current technology zone, meaning that the detection of an Earth-like planet around these stars, by means of direct imaging of the reflected light, is not yet feasible. One problem is that an Earth sibling around most M dwarfs would be located at a planet to star separation well below the inner working angle (IWA) of current coronagraphic instruments, i.e., below the detectable angular separation limit. Another issue is that the combination of stellar luminosity and  simulated Earth luminosity places these planets below achievable planet to star contrast levels. 

Targeting an eclipsing binary has an advantage over a regular binary due to the increased planet to star contrast at the moment of eclipse. This is particularly relevant for planets observable in reflected light, as their brightness would arise from both stars, while the observed starlight would be only from the fainter component of the system. In other words, the eclipse-induced luminosity dip affects only the observed luminosity of the binary system, whereas the reflected luminosity of the planet remains unaltered. 

This paper is structured so that we present a catalog of eclipsing binaries, along with the photometric, astrometric, and spectroscopic properties compiled for this study, in Section~\ref{sec:ebinary_sample}. We consider the detectability of both self-luminous planets and of reflected light Jupiter-, Venus-, and Earth-like planets in Sections~\ref{sec:SLplanets} and \ref{sec:RLplanets}, respectively. We then describe the advantages of our approach with respect to single stars and binary systems that do not eclipse in Section~\ref{sec:advantages} and report our conclusions in Section~\ref{sec:conclusion}.
\section{Eclipsing Binary Sample} \label{sec:ebinary_sample}

We compile a list of eclipsing binary systems from the Catalog of Algol Type Binary Stars \citep{2004AA...417..263B} and the Catalog of Eclipsing Variables \citep{2006AA...446..785M}. To determine the best targets for directly imaging exoplanets during eclipse, we consider those binaries for which the depth of the primary minimum (Dmag) is larger than 2.5, i.e., for which the luminosity dimming factor is at least 10. Overall, 289 eclipsing binary systems satisfy the Dmag constraint: 58 from the first catalog and 231 from the second (also considering the latter's updated version in \citealt{2013AN....334..860A}). In Table~\ref{tab:EBlist} in Appendix~\ref{sec:appendix}, we list all these eclipsing binary systems, along with the photometric, astrometric, and spectroscopic properties relevant for our work here. If these two catalogs report significantly different values of Dmag in the same photometric filter, we exclude that binary from our subsequent analysis, but report it for completeness in Table~\ref{tab:EBlist}.

The Dmag $>2.5$ criterion selects mostly classical Algols, i.e., with an evolutionary class of SA \citep{2013AN....334..860A}. These binaries have a B- or A-type main sequence accretor and a G- or K-type subgiant or giant donor that is large enough to completely eclipse the primary. Observationally, classical Algols are characterized by a deep primary eclipse, shallow secondary eclipse, and ellipsoidal modulations between eclipses due the giant filling its Roche lobe and having a distorted (non-spherical) shape \citep{2004AA...417..263B,2015ApJ...801..113M}.

\subsection{Distances}\label{sec:dist}
Distance is a key factor in limiting direct planet imaging. Because distance data in both catalogs are incomplete, we retrieve parallaxes from the \textit{Gaia} Second Data Release (DR2) \citep{2016A&A...595A...1G,2018A&A...616A...1G}. As explained in \citet{2018A&A...616A...1G}, the parallaxes might be associated with either the photocenter of the system or one of the two components, because all \textit{Gaia} DR2 targets were treated as individual sources.

\begin{figure}[t]
    \includegraphics[width=\columnwidth]{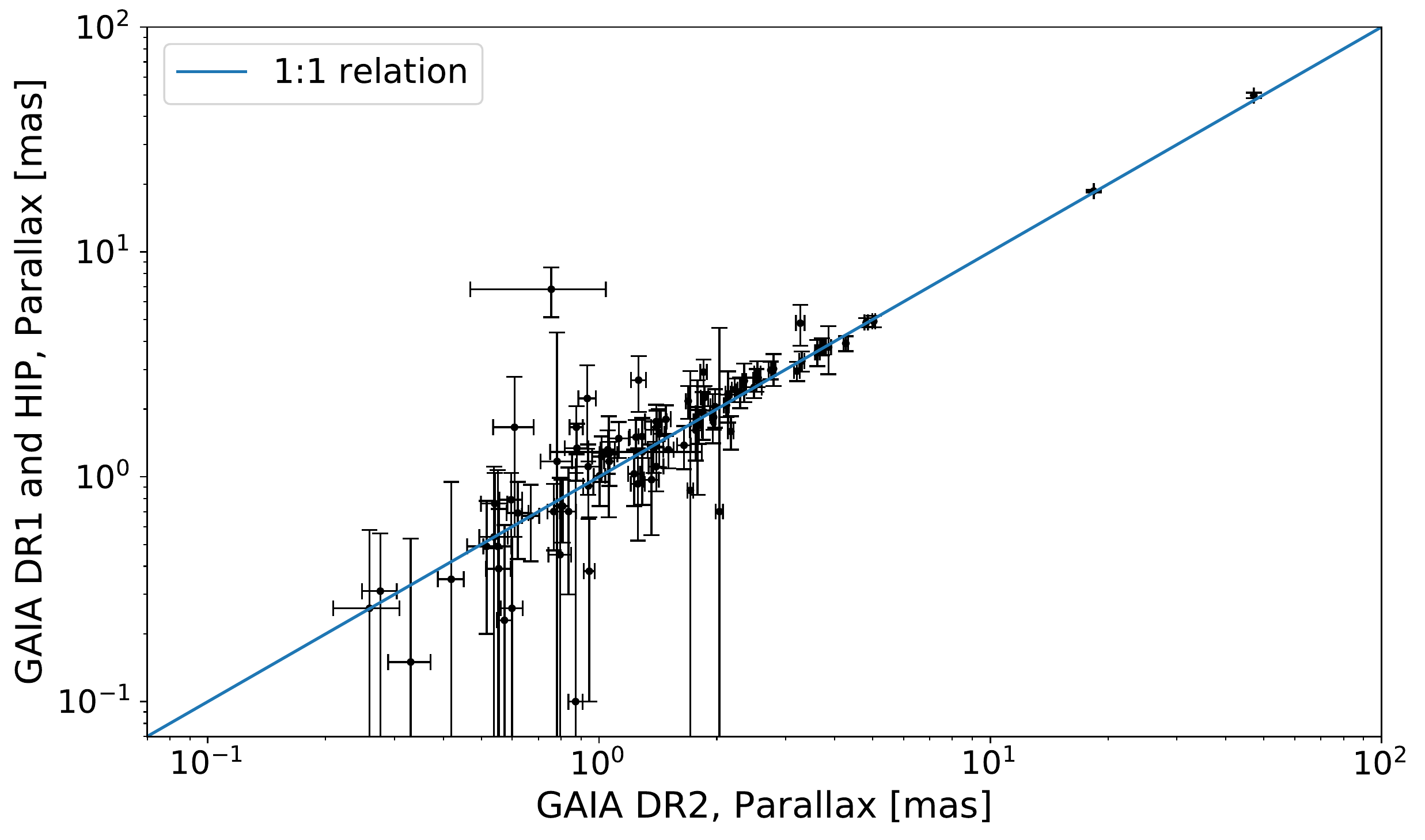}
    \caption{Comparison of parallaxes reported in \textit{Gaia} DR2 (and used in our analysis; Table~\ref{tab:EBlist}) with those in \textit{Gaia} DR1 or \textit{Hipparcos}, where available. The two datasets are consistent to within the published measurement uncertainties and do not show any large systematics.} \label{fig:dist_comparison}
\end{figure}

To check that these DR2 parallaxes are generally consistent with previous measurements, we compare them with \textit{Gaia} DR1 or \textit{Hipparcos} measurements (Figure~\ref{fig:dist_comparison}). The RMS of the residuals with respect to the 1-to-1 line is 0.81 mas, roughly consistent with the mean of the plotted DR1 and \textit{Hipparcos} measurement errors (0.5 mas) and about $10\times$ larger than the mean of the DR2 errors (0.05 mas). There is a slight and expected increase in the scatter at small parallaxes, but, overall, the two datasets are consistent within the published measurement uncertainties. There are no large systematics.

For binary systems characterized by an orbital period longer than 2 yr, there might be a mismatch between the parallaxes or proper motions listed in \textit{Gaia} DR2 with respect to the Tycho-Gaia astrometric solution (TGAS) subset of \textit{Gaia} DR1 \citep{2016A&A...595A...4L}. In our case, the CI Cyg, AR PAV, V381 Sco, and V1329 Cyg systems have periods exceeding this threshold. Considering this and their large distances, we exclude them from our subsequent analysis.

\subsection{Luminosities}\label{sec:lum}
Along with astrometry, \textit{Gaia} DR2 provides stellar luminosities \citep{2018A&A...616A...8A}. The luminosities are inferred via the FLAME module, which is part of the Apsis data processing pipeline \citep{2013A&A...559A..74B}. As the authors specify, there are two potential sources of systematic errors: the adopted bolometric correction BC$_{\text{G},\odot}$ and extinction. The former is estimated to be +0.06 mag. The latter is assumed to be zero when calculating the absolute magnitude, therefore resulting in underestimated luminosity values \citep{2018A&A...616A...8A}.

\begin{figure}[t]
    \includegraphics[width=\columnwidth]{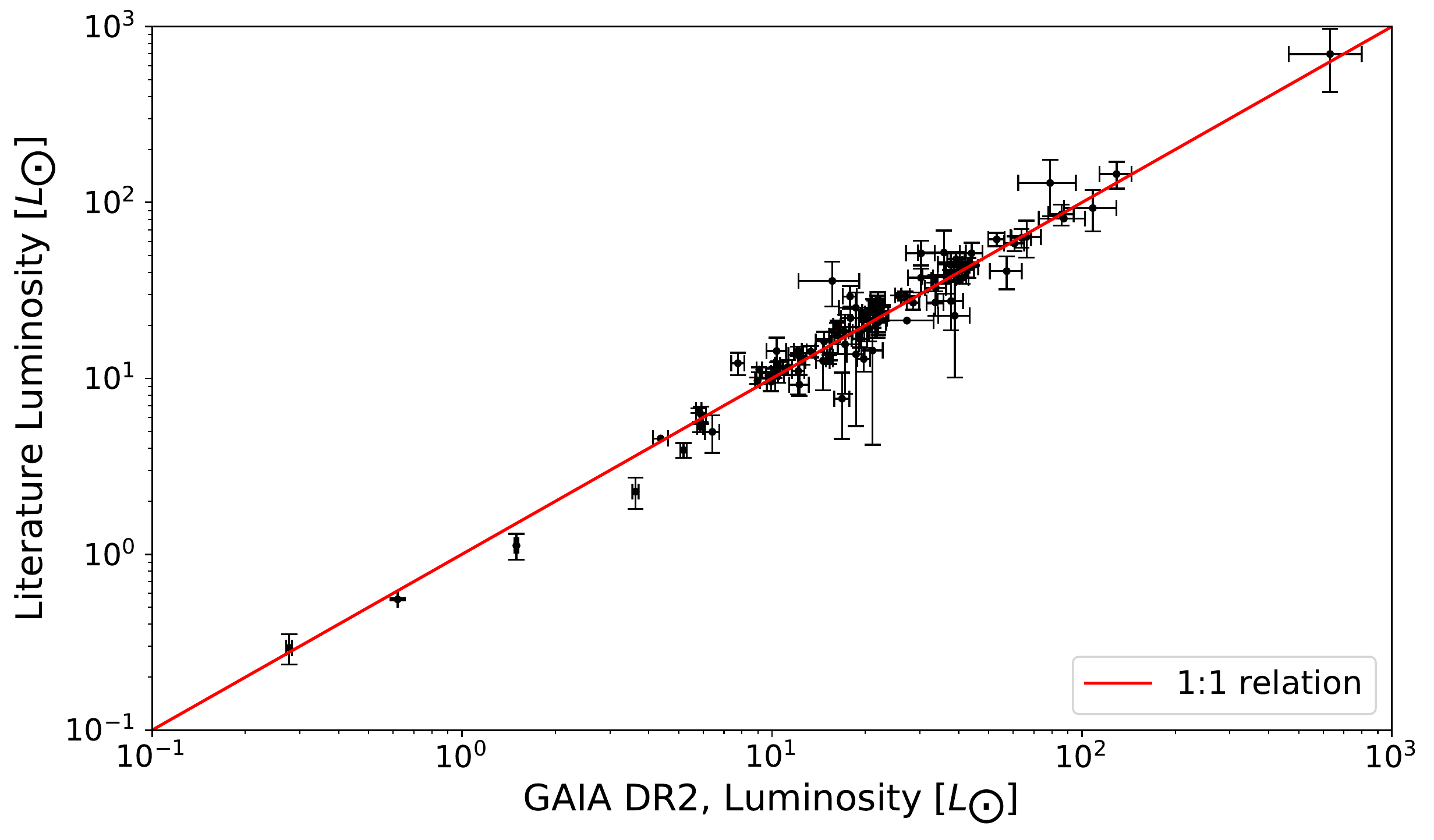}
    \caption{Comparison of bolometric luminosities reported in \textit{Gaia} DR2, and listed in Table~\ref{tab:EBlist}, with those inferred from B and V magnitudes in SIMBAD. The two datasets are consistent within the measurement uncertainties, and there is no evident sign of systematics.}
\label{fig:lum_comparison}
\end{figure}

As we did for the distances in the previous section, we test for
consistency of the luminosities with previous measurements (Figure~\ref{fig:lum_comparison}). The comparison is performed between DR2 luminosities and those inferred from B and V magnitudes from the literature in SIMBAD. Most of the SIMBAD values belong to the \textit{Tycho-2} catalog of the 2.5 million brightest stars \citep{2000A&A...355L..27H}, whereas other entries are taken from the Fourth US Naval Observatory CCD Astrograph Catalog \citep{2013AJ....145...44Z} and the fourth RAVE Data Release \citep{2014AJ....148...81M}. 

Both the B and V magnitudes are used to determine the bolometric correction with the following empirical calibration obtained from the catalog of nearby ($< 8$ pc) stars \citep{1995AJ....110.1838R}:
\begin{equation}
\text{BC}=
\begin{cases}
(B-V) < 1.2: \\
\ \ -0.121112+0.634846(B-V)-1.01318(B-V)^2+\\ 
\ \ 0.125024(B-V)^3;\\
(B-V) > 1.0:\\
\ \ -43.9614+115.958(B-V)-110.511(B-V)^2+\\
\ \ 44.7847(B-V)^3-6.74903(B-V)^4.
\label{eq:Guyon_bolcorrection}
\end{cases}
\end{equation}

The V magnitude is converted from apparent to absolute scale using the distances retrieved from \textit{Gaia} DR2. We then convert the absolute magnitude ($M_V$) into bolometric luminosity with

\begin{equation}
\text{L}_{\text{bol}}=2.512^{-(M_V-4.83)+(\text{BC}-\text{BC}_{\odot})},
\label{eq:Guyon_bolluminosity}
\end{equation}
where BC$_{\odot}=-0.076$.

Considering Figure \ref{fig:lum_comparison}, the RMS of the residuals with respect to the 1-to-1 line is 10.4 L$_{\odot}$, which is roughly consistent with the mean of the plotted SIMBAD \citep{2000A&AS..143....9W} measurement errors (7.3 L$_{\odot}$) and $> 2$ times larger than that  of the \textit{Gaia} DR2 errors (4.1 L$_{\odot}$). There are no obvious systematics.  An additional source of uncertainty in the luminosity would arise if the measurements were made during the eclipse. However, the consistency of the two datasets here suggests that this possibility is unlikely.
\section{Detecting Self-Luminous Planets}\label{sec:SLplanets}

To quantify the advantage of using an eclipsing binary to directly image an orbiting SL planet requires that we estimate the binary's age and assume that the binary and SL planet formed at the same time. We can then model the fading of the SL planet as it cools \citep[e.g.][]{2007ApJ...655..541M,2013A&A...558A.113M,2017A&A...608A..72M} and ask if it is currently bright enough to be detected \citep{2010exop.book..397F} by existing or near-future instruments. In the following discussion, we constrain the ages of our binaries using Algol models \citep{2010AIPC.1314...45V,2017A&A...599A..84M} in Section~\ref{sec:SLages}. Some with large total masses are likely young enough to host detectable SL planets. We then select two with the youngest age ranges as potential targets for SL planet direct imaging follow-up in Section~\ref{sec:SLbest}.
\subsection{Ages}\label{sec:SLages}

Neither the eclipsing binary catalogs from which we construct our sample nor the \textit{Gaia} DR2 report ages, which are expected with the third release of \textit{Gaia}. Determining the ages of our binaries is challenging regardless; as discussed earlier, our Dmag $>2.5$ criterion tends to select classical Algols. Late B-dwarf classical Algols are field blue stragglers, and, because they are rejuvenated by the mass transfer, are older than would be expected from the main sequence lifetime of the current primary \citep{1971ARA&A...9..183P,1984A&A...140..373G,1987ApJ...313..727I}. The total mass of the binary system is a better proxy for age.

Mass transfer is nearly conservative for binaries with initially A- or late-B type primaries that interact via Case A Roche-lobe overflow \citep{2010AIPC.1314...45V,2017A&A...599A..84M}. Under these circumstances, the total mass of the current system ($M_{\text{tot}}$) is assumed equal to the sum of the components' masses when they reached the zero-age main sequence. Then, the initial mass of the primary is 
\begin{equation}
M_{\text{primary}}=\frac{M_{\text{tot}}}{q+1},
\label{eq:primarymass}
\end{equation}
where $q$ is the initial mass ratio. If the components are initially the same, $q = 1$ and the minimum $M_{\text{primary}}$ is $0.5 M_{\text{tot}}$. For stable mass transfer to occur with a late-B primary, $q \ge 0.4$ and the maximum $M_{\text{primary}}$ is $0.7 M_{\text{tot}}$. Given that the initial primary has now evolved into a subgiant or giant, and that the time it spent on the main sequence was much longer, its main sequence lifetime provides an estimate of the binary's age. Assuming $0.5 M_{\text{tot}} \leq M_{\text{\text{primary}}} \leq 0.7 M_{\text{tot}}$, with $M_{\text{tot}}$ obtained from the component masses in \citet{2004AA...417..263B}, and the relationship between stellar mass and main sequence lifetime, we convert this $M_{\text{primary}}$ range into an age range for each of our binaries. This range could extend to younger ages if mass loss affects the transfer process ($M_{\text{primary}}>0.7M_{\text{tot}}$), but we proceed with the conservative (older) age estimates above.

\begin{table*}[t]
\begin{center}
\caption{Best Eclipsing Binary Targets for Direct Detection of Self-Luminous Exoplanets \label{tab:bestSL}}
\begin{tabular}{ccccccccccccc}
\hline 
\hline 
Name & $m_V$ & Dmag$_V$ & $m_J$ & Dmag$_J$ & d & Period & $t_{\text{Dmag}}$ & $a_{\text{bin}}$ & $M_1$ & $M_2$ & Spec Type & Age Interval\\
 & & & & & [pc] & [days] & [min] & [AU] & [M$_{\odot}$] & [M$_{\odot}$] &  & [Myr]\\
\decimalcolnumbers
U Cep & 6.9 & 2.54 & 6.47 & 1.10 & 198.6 & 2.5 & 90 & 0.07 & 4.20 & 2.30 & B7V+[G8III-IV] & 215-525 \\
AC Sct & 10.0 & 2.60 & 9.72 & 1.62 & 985.6 & 4.8 & 168 & 0.11 & 2.80 & 1.40 & B9+[G0IV] & 640-1365 \\
\end{tabular}
\end{center}
\tablecomments{(1) General Catalog of Variable Stars designation, (2) Magnitude at maximum brightness in $V$-band, (3) Depth of primary minimum in $V$-band, (4) Magnitude at maximum brightness in $J$-band, (5) Inferred depth of primary minimum in $J$-band, (6) Distance inferred from parallax, (7) Binary period, (8) Duration of totality in primary eclipse, (9) Projected separation between binary components, (10) Mass of primary component, (11) Mass of secondary component, (12) Spectral type, (13) Binary age estimated from total binary mass $M_{tot} = M_1 + M_2$. Columns (2), (3), (10), (11), and (12) are extracted from \citet{2004AA...417..263B}, column (9) from \citet{1980AcA....30..501B}, column (4) from 2MASS Catalog \citep{2003yCat.2246....0C}, column (5) from Eq.~\ref{eq:JDmag}, column (6) from \textit{Gaia} DR2, columns (7) and (8) from \citet{2013AN....334..860A}. The values in column (13) are inferred from $M_{tot}$ and Eq.~\ref{eq:primarymass}, assuming $q=0.4$ to 1. The evolutionary class of both targets is SA, i.e., classical Algols \citep{2013AN....334..860A}.}
\end{table*}

At present, there are few detections of SL planets around binaries with massive stars. Observational studies such as the on-going BEAST survey \citep{2019A&A...626A..99J} aim to address the question of exoplanets around B-type stars. Our binaries have the potential to shed more light on whether the incidence of massive planets increases or declines with host stellar mass, thus constraining the stellar mass interval within which planet formation is favorable \citep{2011ApJ...736...89J,2019A&A...626A..99J}.
\subsection{Best targets}\label{sec:SLbest}

We shortlist the best eclipsing binaries for observational follow-up according to the criteria discussed previously: 1) depth of primary eclipse larger than 2.5 mag, 2) accessibility with current or near-future technology, and 3) likelihood that the binary is young, and thus that any orbiting SL planet is luminous, based on the binary total mass. The two youngest eclipsing binaries satisfying these criteria are U Cep and AC Sct, with age intervals of 215-525 Myr and 640-1365 Myr, respectively (Table~\ref{tab:bestSL}).

An additional consideration for selecting suitable targets is whether the binary system has a tertiary companion. \citet{2006A&A...450..681T} show that short period ($P$) binaries are more likely to have a third component which, if orbiting at small separation, could suppress planet formation \citep{2019arXiv191201699M}. In particular, no circumbinary planet has been found around a $P<7$ days binary, which may be due to the presence of a tertiary \citep{2016MNRAS.455.3180H}. In our case, U Cep ($P=2.5$ days) is known to have a third companion \citep{2018ApJS..235....6T} at $\sim$2800 AU, so its influence on a potential planet's dynamical stability is negligible. It is not known whether AC Sct ($P=4.8$ days) has a tertiary component.

The projected separation between the binary stellar components is 0.07 AU for U Cep and 0.11 AU for AC Sct \citep{1980AcA....30..501B}. Therefore, we would expect any potential planet to lie on a P-type orbit. Because the eccentricities of the systems are not available, we cannot assess the long-term stability of the planetary orbits \citep{1999AJ....117..621H,2016AJ....151..111Q} at this time.

We add two more potential targets, V621 Cen and RW Mon,
if we relax our deep eclipse criterion from Dmag $>2.5$ to $>2.0$ mag, a contrast improvement of $\sim$6$\times$. The total mass of V621 Cen corresponds to an age range of 196-478 Myr, i.e., comparable to U Cep's, but detecting SL planets around V621 Cen would require future planned facilities due to its large distance (1.8 kpc). RW Mon is closer (505 pc) and has an age range of 880-2150 Myr. RW Mon's period variations may arise from a close tertiary companion \citep{2011NewA...16..253S}, so targeting this system would not only test our direct imaging method, but also reveal the nature of any third component. There could be other good targets within our sample, but some binaries have missing or conflicting data, e.g., Dmag, which prevents us from evaluating them.

For U Cep and AC Sct, we calculate whether the eclipse would increase the infrared SL planet-star contrast to that required by the instruments. We consider planet masses between 0.5 and $10M_J$, determining the corresponding planet $J$-band (1.25 $\mu m$) magnitude at the age of the binaries with the Sonora evolutionary models (Marley et al. 2020, in prep.). We then estimate the planet-star contrast at eclipse and compare it with the technology regions as in Figure~\ref{fig:Belikov}.

The planet-star contrast during the eclipse is the ratio of the planet luminosity to the primary minimum. The eclipsing binary catalogs report the Dmag value for U Cep and AC Sct in the $V$-band. Thus, we estimate Dmag in $J$-band based on the $V-J$ color of both binary components\footnote{The $V-J$ colors are retrieved from \href{http://www.pas.rochester.edu/~emamajek/spt/}{http://www.pas.rochester.edu/~emamajek/spt/}. For U Cep's primary and secondary, $V-J$ is -0.24 and 1.57, respectively; for AC Sct, this color is -0.09 and 1.06.}. Formally, we write the fluxes normalized to the total flux in the $V$-band, i.e., $F_{\text{p},V}+F_{\text{s},V}=1$, where $F_{\text{p},V}$ and $F_{\text{s},V}=10^{-0.4\text{Dmag}_V}$ are the primary and secondary $V$-band fluxes, respectively. Then, we have
\begin{equation}
\text{Dmag}_J=-2.5\log\left(\frac{F_{\text{s},J}}{F_{\text{tot},J}}\right)
\label{eq:JDmag}
\end{equation}
where $F_{\text{s},J}=F_{\text{s},V}10^{0.4(V-J)_{\text{s}}}$ and $F_{\text{tot},J}=F_{\text{p},J}+F_{\text{s},J}$, where $F_{\text{p},J}=F_{\text{p},V}10^{0.4(V-J)_{\text{p}}}$.

The $J$-band luminosity during the eclipse is then calculated as
\begin{equation}
\text{L}_{J,e}=10^{-0.4\text{Dmag}_J}\text{L}_J
\label{eq:lumecl}
\end{equation}
where L$_J$ and L$_{J,e}$ are the $J$-band luminosities at maximum brightness and primary minimum, respectively.

\begin{figure}[t]
    \centering
    \includegraphics[width=\columnwidth]{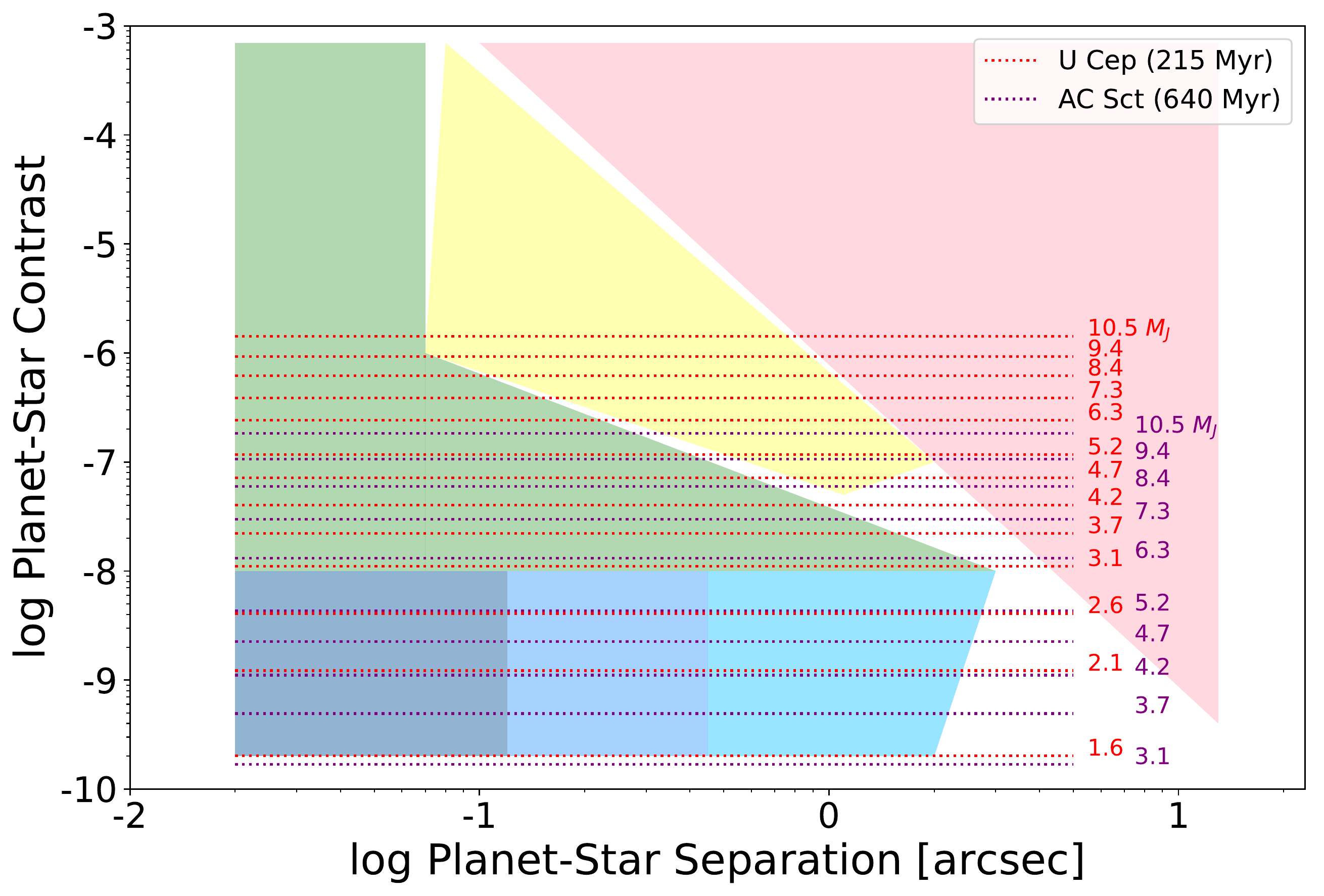}
    \caption{Planet-star $J$-band contrast versus separation for simulated SL planets around U Cep (red) and AC Sct (purple). The planet-star contrast is the ratio of the planet to binary luminosity at eclipse. The numerator is derived from the Sonora evolutionary models (Marley et al. 2020, in prep.), assuming that the planet is as old as the estimated binary age. The denominator is equivalent to the $J$-band Dmag, which is estimated from the $V-J$ colors of the binary components (see Eq.~\ref{eq:JDmag}). Current and future observational facilities are denoted by the colored regions as in Figure~\ref{fig:Belikov}. Around U Cep,  current ground- or near-future space-based (yellow), future ground-based (green), and future space-based (blue) instruments can detect $\gtrsim5M_J$, 3-4$M_J$, and 1.5-2.5$M_J$ SL planets, respectively. Around AC Sct, an older binary, these limits rise to $\gtrsim 9M_J$, 6-8$M_J$, and 3-5$M_J$, respectively.}
   \label{fig:SLmosaic}
\end{figure}

Figure~\ref{fig:SLmosaic} illustrates the planet-star contrast versus separation plane for U Cep and AC Sct. Around U Cep, planets of $\gtrsim4.5M_J$ reach contrast levels $\gtrsim10^{-7}$ and are thus detectable with current ground- or near-future space-based instruments. Planets of 3-4$M_J$ and roughly 1.5-2.5$M_J$ achieve contrast levels associated with future ground- ($\sim$10$^{-8}$) and space-based ($\sim$10$^{-9}$) facilities, respectively. Around AC Sct,  $\gtrsim$9$M_J$ planets can be detected with current ground- or near-future space-based instruments, while 6-8$M_J$ and 3-5$M_J$ planets require future ground- and space-based observatories, respectively.

In Figure~\ref{fig:SLmosaic}, the technology regions are the same as in Figure \ref{fig:Belikov} and defined by the instrument performance and inner-working angle. The location of the simulated planets on the planet-star separation axis is arbitrary. Current ground- or near-future space-based instruments are characterized by observable separations of roughly 0.1-2 arcsec. At the distances of U Cep and AC Sct, this range translates into orbital semi-major axes of 20-400 AU and 100-2000 AU, respectively, consistent with those of known directly imaged planets around binaries \citep{2016MNRAS.460.3598S}.

\begin{figure}[t]
    \centering
    \includegraphics[width=\columnwidth]{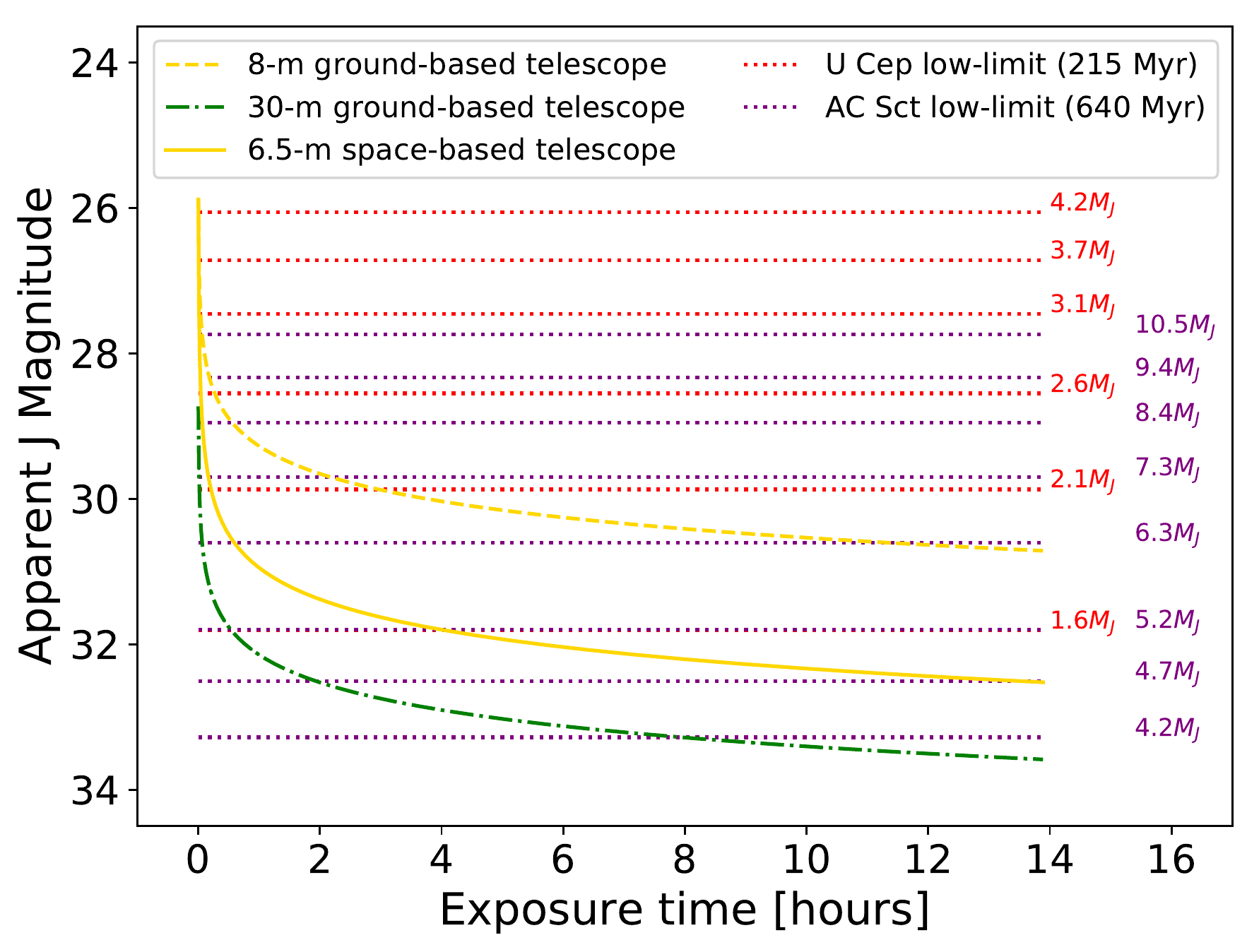}
    \caption{Comparison of simulated SL planet $J$-band
    magnitude with observational detection limits, as a function of exposure time and planet mass. We consider different observational capabilities corresponding to some of the technology regions in Figure \ref{fig:SLmosaic}:
    a current 8-m (yellow dashed line) and a future 30-m (green dot-dashed) ground-based telescope and a near-future 6.5-m (yellow dashed) space facility. We assume photon-limited detections and require a signal-to-noise ratio of 10. For the ground-based observatories, we adopt an overall efficiency of 20\% and a sky background contamination of $16.7\;\text{mag}/\text{arcsec}^2$. For the space-based case, we assume a 30\% efficiency and a zodiacal background contamination of $22.2\;\text{mag}/\text{arcsec}^2$. SL planet magnitudes are calculated at the distances of U Cep (red) and AC Sct (purple). Around U Cep, within a 1 hour exposure, a $\gtrsim4.5M_J$ planet is detectable with current ground-based or near-future space-based instruments, a $\sim$1.5$M_J$ planet with future ground-based facilities. Around AC Sct, these limits rise to $\sim$9$M_J$ and 6-8$M_J$, respectively.}
   \label{fig:Jlimits}
\end{figure}

Are the SL planets in Figure~\ref{fig:SLmosaic} bright enough to be detectable? Figure~\ref{fig:Jlimits} shows the change of the direct imaging detection limits as a function of exposure time and planet mass. For U Cep, in a $<$1 hour exposure, $\gtrsim4.5M_J$ planets are detectable with current ground- or near-future space-based instruments, $\sim$1.5$M_J$ planets with a future ground-based observatory. For AC Sct, these limits are $\sim$9$M_J$ and 6-8$M_J$, respectively.

The eclipse durations and short orbital periods of U Cep and AC Sct allow for efficient scheduling of their observations. Ideally, the duration of totality would exceed the exposure time required for planet detection and the eclipses would be frequent enough to allow for repeat observations during a typical observing run of several nights. As discussed above, a $\sim$1 hour exposure is required to detect planets around U Cep and AC Sct with current ground- or near-future space-based facilities. In comparison, the totality of the primary eclipse for U Cep lasts 90 minutes, during a primary eclipse of 9.0 hours that occurs every 2.5 days. For AC Sct, totality lasts approximately 168 minutes, during a primary eclipse of 16.1 hours every 4.8 days. Therefore, we can achieve the detection limit within a reasonable time frame.

\begin{figure}[t]
    \includegraphics[width=\columnwidth]{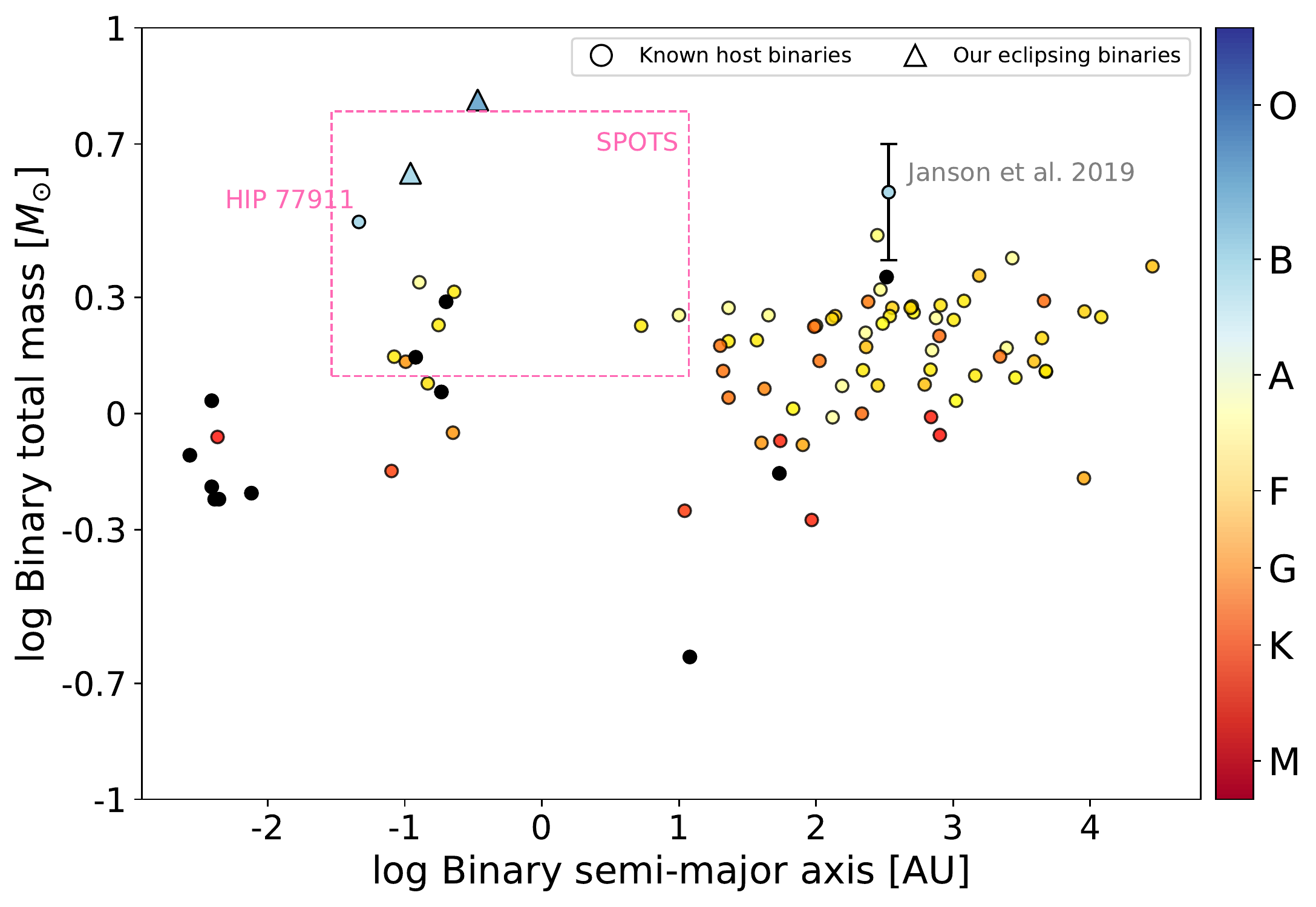}
    \caption{Binary total mass versus semi-major axis plane comparing our two best eclipsing binary targets, U Cep and AC Sct, with binaries known to host planets \citep{2016MNRAS.460.3598S}. The datapoints are color-coded by primary spectral type; black dots are systems whose primary type is not available. As a consequence of our Dmag and binary age selection criteria, U Cep and AC Sct have earlier type primaries and larger total masses than typical for known host binaries. This distinction would expand the parameter space of SL hosting binaries, were  planets discovered in our target systems. We highlight HIP 79098 AB (gray), which has a candidate substellar companion \citep{2019A&A...626A..99J} and HIP 77911 (pink), which has a candidate planetary-mass companion \citep{2018A&A...619A..43A}; these systems have primaries with similar spectral types to our targets, but lower total masses. In comparison to the parameter boundaries of the SPOTS survey \citep{2018A&A...619A..43A} (pink box), our targets fall on the low side of the semi-major axis distribution and the high end of total mass.} 
    \label{fig:star_niche}
\end{figure}

In Figure~\ref{fig:star_niche}, we compare the properties of U Cep and AC Sct  (Table~\ref{tab:bestSL}) with those of binaries known to host exoplanets \citep{2016MNRAS.460.3598S}, regardless of the detection technique. Our eclipsing binary targets have earlier type primaries and larger total masses due to the Dmag and age criteria applied above. Specifically, among the 97 binaries known to host planets and whose primaries have been spectrally classified \citep{2016MNRAS.460.3598S}, there are none with B- or A-type primaries or with total masses above 4$M_{\odot}$ like our targets. Thus, targeting our binaries would expand the spectral and total mass ranges of host binaries, were planets discovered there.

\section{Searching for Reflected Light Planets} \label{sec:RLplanets}

Of the few direct imaging detections around binaries so far, none has included a RL planet. Therefore, it is interesting to test whether our approach would give access to this unexplored territory. There is an added benefit to using eclipsing binaries for RL planets relative to SL ones. At eclipse, the  planet-star contrast is improved, while the RL planet is still brightened by the light of \emph{both} binary components. RL planets have evolved and cooled, so, unlike for SL planets, there is no age constraint for shortlisting targets here.

\begin{table*}[t]
\begin{center}
\caption{Best Eclipsing Binary Targets for Direct Detection of Reflected Light Exoplanets \label{tab:RLbest}}
\begin{tabular}{ccccccccc}
\hline 
\hline 
Name & $m$ & Dmag & Filter & $m_V$ & d & Period & t$_{\text{Dmag}}$ & Spec Type\\
 & & & & & [pc] & [days] & [min] &\\
\decimalcolnumbers
RR Cae & 14.88 & 3.30 & $B$ & 14.40 & 21.21 & 0.3 & 11 & WD+M5-6V\\
V1412 Aql & 15.67 & 2.63 & $V$ & 15.67 & 22.93 & \nodata & \nodata & DC7\\
RT Pic & 9.90 & 2.60 & $p$ & 9.07 & 54.39 & \nodata & \nodata & G8V\\
\end{tabular}
\end{center}
\tablecomments{(1) General Catalog of Variable Stars designation, (2) Magnitude at maximum brightness, (3) Depth of primary minimum, (4) Filter band for $m$ and Dmag, (5) Magnitude at maximum brightness in $V$-band, (6) Distance, (7) Binary period, (8) Duration of totality in primary eclipse, (9) Spectral type. Columns (2), (3), (4), (7), (8), and (9) are extracted from \citet{2013AN....334..860A}, column (6) is inferred from \textit{Gaia} DR2 parallax, and column (5) is taken from SIMBAD. RR Cae's evolutionary class is DW, i.e., white dwarf system \citet{2013AN....334..860A}.}
\end{table*}

\begin{figure*}[t]
    \centering
    \includegraphics[width=0.75\textwidth]{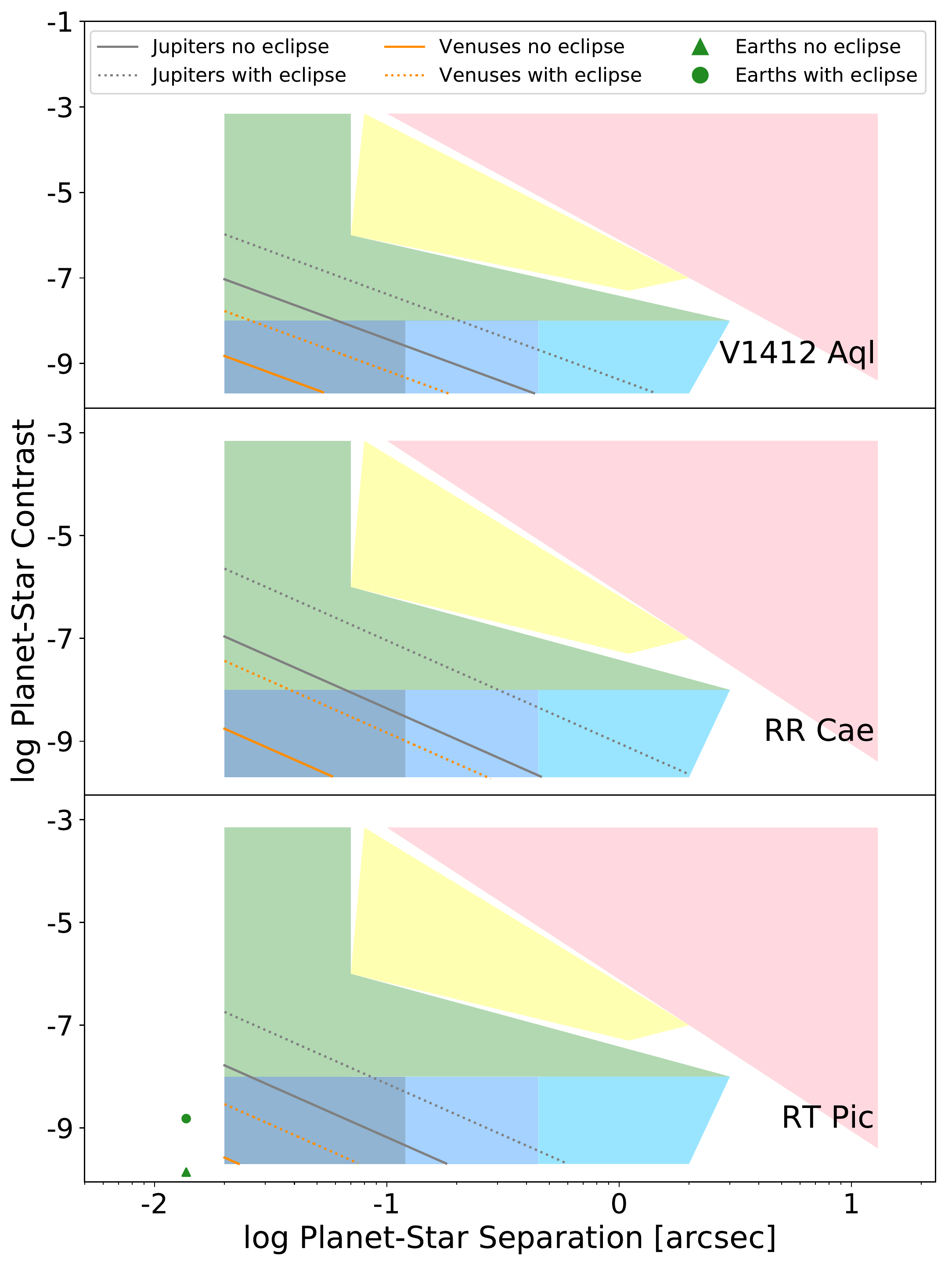}
    \caption{Testing the detectability of RL planets during the eclipse of the binary. Once again, we plot the observational space from Figure~\ref{fig:Belikov}, but now for the closest eclipsing binaries in our sample. We consider simulated Jupiter- (gray) and Venus-like (orange) planets at increasing separations from the hosts, starting from 20 mas, the detectable angular separation limit of near-future facilities (green and dark blue regions). We plot the $V$-band contrasts both during (dotted lines) and before/after the eclipse (solid lines). The planet flux observed would decline from left to right. A habitable (i.e., flux received from the binary equal to the solar constant) Earth-like (same radius and albedo) planet only appears on this plot for RT Pic, at eclipse (green circle) and at other times (green triangle). The planet-star contrasts and separations of ``Jupiters'' and ``Venuses'' orbiting our closest binaries are accessible with future ground- and space-based coronagraphs. While the contrast of the habitable Earth-like planet around RT Pic is achievable with planned space-based instruments, the planet-star separation is too small.
    \label{fig:RLmosaic}}
\end{figure*}

In selecting potential RL targets, the main observational limitation is the distance of the binary from us. Given that a typical inner working angle of current coronagraphs (yellow region; Figure~\ref{fig:RLmosaic}) is on the order of 100 mas, we do not expect to observe RL planets on tight orbits in faraway systems. At 50 pc away, the separation corresponding to a planet on a 1 AU orbit is only 20 mas, whereas an orbit of 5 AU or larger is observable. Therefore, we consider only the three deep (Dmag $\gtrsim$ 2.5) eclipsing binaries within roughly 50 pc: V1412 Aql at 22.9 pc, RR Cae at 21.2 pc, and RT Pic at 54.4 pc (Table~\ref{tab:RLbest}). In the case of RR Cae, a $4.2\;M_J$ planet has already been discovered via the eclipse timing method at $5\;\text{AU}$ \citep{2012MNRAS.422L..24Q}.

For RR Cae (11 min totality, $\sim$14 min eclipse duration, 7.2 hour orbital period), there is little time during totality for observations, but we can build up a long exposure by observing for a short fraction of each night over multiple nights. For the other two targets, the totality, eclipse duration, and binary period are not known at present.

To determine whether an RL planet is detectable for these binaries, we build the observational space in Figure~\ref{fig:RLmosaic} as in Figure~\ref{fig:Belikov}. We assume that Jupiter- and Venus-like planets (i.e., with the same sizes and albedos as the originals) have observed planet-star separations of at least 20 mas, the detectable angular separation limit of future technology (green and dark blue regions in the figure). We implicitly assume that these simulated planets could lie this physically close to their binary. The separation between the binary components is not available in the literature; therefore, we cannot predict whether the planets lie on a P- or S-type orbit.

For each binary, we derive the planet-star contrast at maximum binary brightness (L$_{\text{pl}}/$L$_V$) and during the eclipse (L$_{\text{pl}}/$L$_{V,e}$) in the $V$-band. The luminosity at eclipse (L$_{V,e}$) is estimated from Eq.~\ref{eq:lumecl} using the $V$-band luminosity outside eclipse (L$_V$) and the observed Dmag, which is reported in other photometric filters for RR Cae ($B$) and RT Pic ($p$) (see Table~\ref{tab:RLbest}). Unlike in Section~\ref{sec:SLbest}, we do not have enough information here about the primary and secondary (i.e., spectral type and/or color), and so cannot convert Dmag to a common band. Given that Dmag may decrease with increasing wavelength, the plotted contrast boost could be overestimated for RR Cae and RT Pic.

The luminosity of the planet due to reflection (L$_{\text{pl}}$) is obtained from the definition of the albedo. Assuming the planet to be a disk of radius R$_{\text{pl}}$ and manipulating the ratio between the incident and reflected flux yields
\begin{equation}
\text{L}_{\text{pl}}=\text{L}_{\text{tot}}\frac{\text{R}_{\text{pl}}^2}{a^2}\frac{\alpha}{4},
\label{eq:Rlight}
\end{equation}
where $a$ and $\alpha$ are the semi-major axis of the planetary orbit and the albedo, respectively. The values of $a$, $\alpha$, and R$_{\text{pl}}$ are taken from the NASA fact sheets for each planet\footnote{For Jupiter, Venus, and Earth:
\href{https://nssdc.gsfc.nasa.gov/planetary/factsheet/jupiterfact.html}{/jupiterfact.html}, \href{https://nssdc.gsfc.nasa.gov/planetary/factsheet/venusfact.html}{/venusfact.html}, and \href{https://nssdc.gsfc.nasa.gov/planetary/factsheet/earthfact.html}{/earthfact.html}, respectively.}.

Figure~\ref{fig:RLmosaic} shows that for Jupiter- and Venus-like planets with planet-star separations within the technology limits, the planet-star contrasts also will be accessible.

\begin{figure}[t]
    \includegraphics[width=\columnwidth]{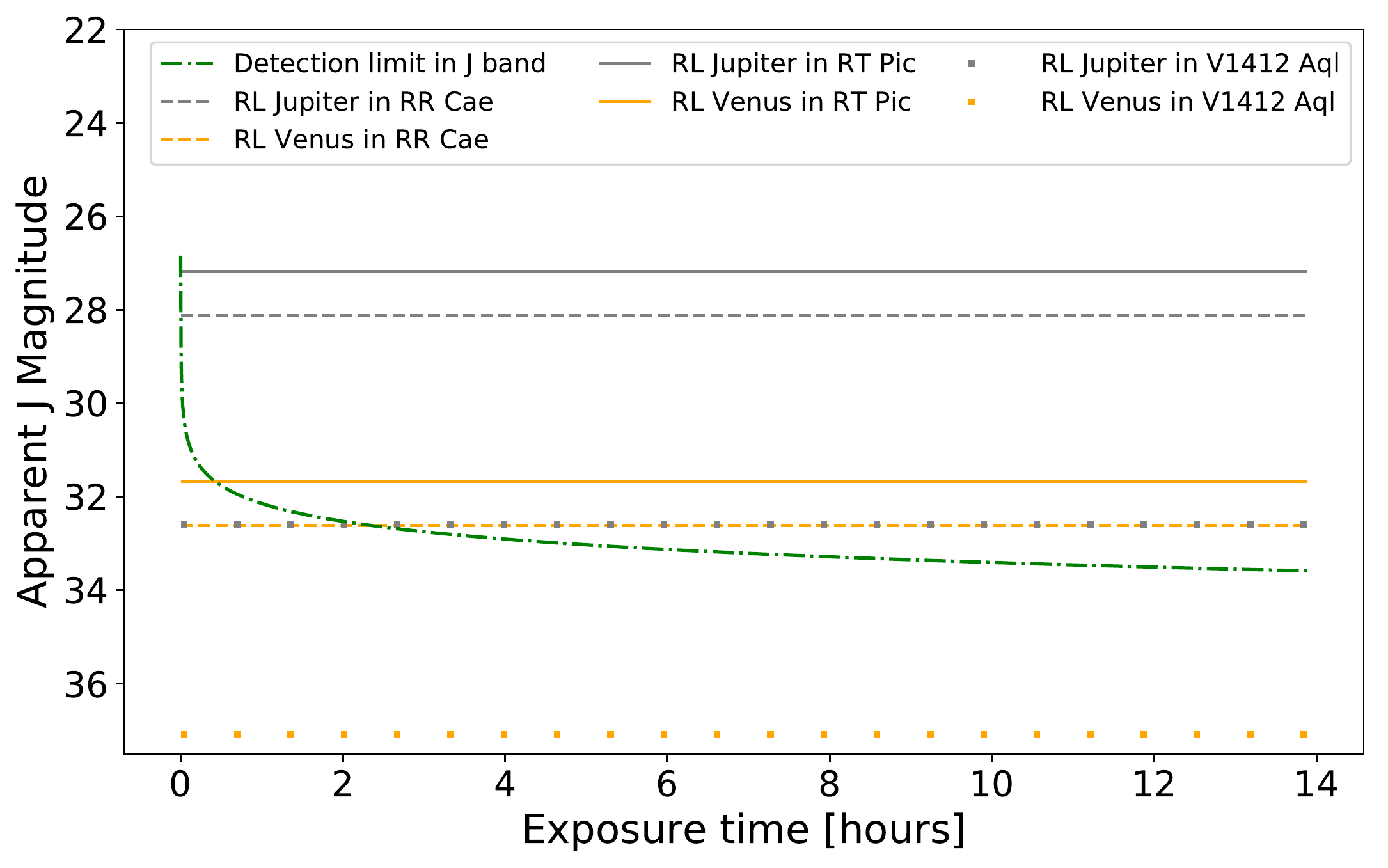}
    \caption{Comparison of simulated $J$-band magnitude with observational detection limit, as a function of exposure time. We assume a 30-m ground-based telescope with the same parameters as in Figure~\ref{fig:Jlimits} (green dot-dashed line). Also plotted are the apparent magnitudes of a ``Jupiter'' and ``Venus'' at a separation of 20 mas
    from V1412 Aql (dotted gray and dotted orange lines, respectively), RR Cae (dashed gray and dashed orange lines), and RT Pic (solid gray and solid orange lines). RT Pic is the most promising target given the higher brightness of the planets there. A Jupiter-like planet around all three binaries and a Venus-like planet around RR Cae and RT Pic would be detected in two hours.
    \label{fig:groundRL}}
\end{figure}

\begin{figure}[t]
    \includegraphics[width=\columnwidth]{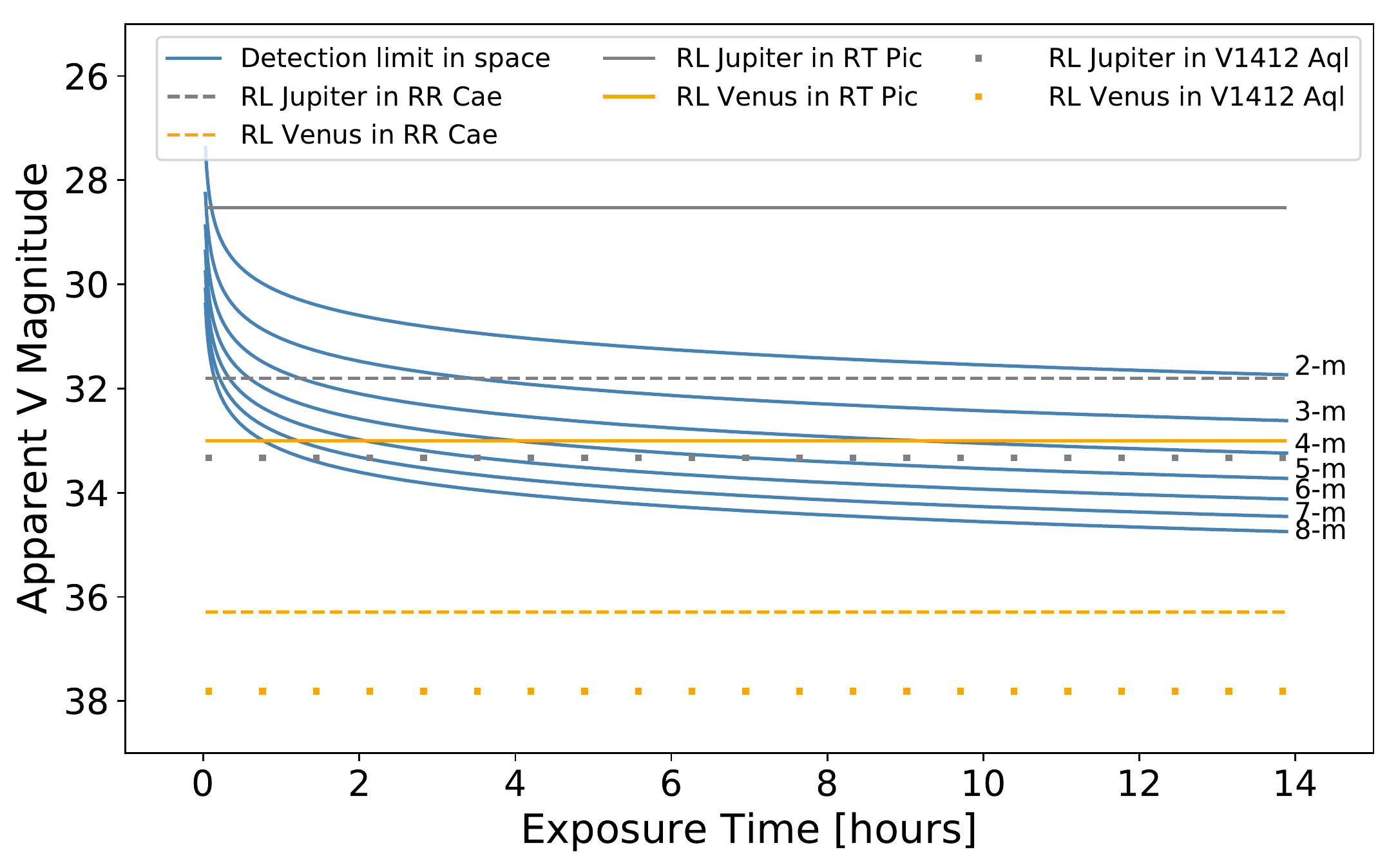}
    \caption{Comparison of simulated $V$-band magnitude with observational detection limit, as a function of exposure time, for the Venus- and Jupiter-like planets in Figure~\ref{fig:groundRL}, assuming future space-based telescopes of different apertures and same parameters as the space-based case in Figure~\ref{fig:Jlimits} (blue solid lines). The Venus-like planet (orange) around RT Pic would be detected in the $V$-band in less than 10 hours with a 4-m telescope. A Jupiter-like planet (gray) around RT Pic would be observable in less than 1 hour with a 2-m aperture, in $\sim$4 hours around RR Cae with a 3-m aperture, and in $\sim$7 hours around V1412 Aql with a 5-m aperture.
    \label{fig:spaceRL}}
\end{figure}

To further quantify the detectability of these simulated RL planets, we estimate the required exposure times for future ground- (Figure~\ref{fig:groundRL}) and space-based (Figure~\ref{fig:spaceRL}) observations, similarly to Figure~\ref{fig:SLmosaic}. For ground-based and space-based imaging, respectively, we assume that the observations are carried out in the $J$- (1.25 $\mu m$) and $V$-band (555 nm) 
to achieve the best planet-star contrast and separation. For the ground-based case, a large telescope aperture ($\sim$30 m) with extreme-AO in the near-IR leads to the best contrast, given that extreme-AO in visible light is considerably more challenging.

For a Jupiter-like planet at 20 mas separation, the $J$-band magnitude is brighter than 33 mag around all three eclipsing binaries, implying a ground-based detection within 2 hours (Figure~\ref{fig:groundRL}). With a space-based telescope (Figure~\ref{fig:spaceRL}), a Jupiter around RT Pic would be detected in less than 1 hour with a 2-m aperture, in $\sim$4 hours around RR Cae with a 3-m aperture, and in $\sim$7 hours around V1412 Aql with a 5-m aperture. For a Venus-like planet at 20 mas separation, the $J$-band magnitude would be 37.1, 32.6, and 31.7 mag, respectively, for V1412 Aql, RR Cae, and RT Pic, requiring at most two hours to reach the detection limit around RR Cae and RT Pic with a ground-based telescope. With a space-based telescope, the $V$-band detection limit of a Venus would be accessible in less than 10 hours, even with a 4-m aperture, but only around RT Pic. 

To Figure~\ref{fig:RLmosaic}, we add an Earth-like (same radius and albedo) planet, assuming that it is habitable, i.e., that the bolometric flux it receives from the binary is equal to the solar constant. Therefore, the semi-major axis of the planet is obtained from $\sqrt{\text{L}_{tot}}$ in AU. We do not consider the eccentricity of the binary or the gravitational interaction as the secondary moves in its orbit, both of which alter the habitable zone boundaries over time \citep{2013ApJ...777..166H,2013ApJ...777..165K,2014MNRAS.443..260J}. 

While the ``Earth'' in RT Pic lies at a planet-star contrast detectable by space-based telescopes planned for the 2030's, its separation from the binary is beyond the capability of those instruments. Indeed, the planet-star separation would be 13 mas, the planet-star $V$-band contrast $1.5\times10^{-9}$ during the eclipse, and the planet $J$-band magnitude 32.4 and $V$-band magnitude 33.7. The detection limit in the $J$-band would be achievable in $\sim$2 hours with a ground-based telescope and in the $V$-band in $<$7 hours with a 6- to 8-m space-based telescope. For the ground-based case, the $1.5\times10^{-9}$ planet-star contrast is not achievable. For the space-based telescopes, the diffraction limits are 23.3 mas ($\lambda$ = 555 nm, D = 6 m) and 17.5 mas ($\lambda$ = 555 nm, D = 8 m). Detection at 13 mas would require that the $1.5\times10^{-9}$ planet-star contrast be achieved at 0.56 $\lambda$/D and 0.74 $\lambda$/D, respectively. Coronagraphs currently deliver deep contrast levels at $\gtrsim$2 $\lambda$/D \citep{2006ApJS..167...81G}, so detecting an Earth-like planet around an eclipsing binary will require larger apertures and/or further advances in coronagraph technologies.\\
\section{Advantages of our method and targets} \label{sec:advantages}

As discussed in the previous section, for reflected light planets, the eclipse improves the observed planet-star contrast by dimming the primary while the planet remains illuminated by both stars. Observing eclipsing binaries has several other advantages for directly imaging both SL and RL planets: 1) The reduction of the binary to a point-like source during eclipse makes coronagraphy feasible. 2) The increase in planet-star contrast during eclipse makes fainter planets accessible. 3) The contrast boost allows detection of planets in intrinsically brighter, and thus more massive, stellar systems. 

Eclipsing binaries are observable with a coronagraph, which would block the point-like light of the superimposed stellar components during the eclipse. In this way, our method incorporates both a natural starshade (the eclipse) and a man-made coronagraphic measurement. Light leakage during the coronagraphic measurement would be minimized, because it is possible to predict the time and duration of the binary eclipse accurately. In fact, one of our eclipsing binaries, RR Cae, has been successfully targeted in an exoplanet search with the eclipse timing variation technique \citep{2012MNRAS.422L..24Q}, for which timing accuracies on the order of seconds are required \citep{2010MNRAS.405..657S}

Our method would work best during the total eclipse of the primary, when the binary is point-like and darkest. However, we could still observe at partial eclipse, provided that the angular separation between the binary components is below the angular stellar size tolerance of the coronagraph (on the order of 0.1 $\lambda$/D). For our best SL targets (U Cep and AC Sct), the projected binary separation is $\sim$0.01 $\lambda$/D in the $J$-band with a 30-m telescope. Thus, observations during partial eclipse are potentially useful. For our best RL targets (V1412 Aql, RR Cae, and RT Pic), we lack sufficient information to infer the binary separations and make a similar evaluation. 

The performance of a coronagraph with a small IWA ($<$3 $\lambda$/D) degrades with increasing stellar angular size \citep{2006ApJS..167...81G}, even if the source size is well below the diffraction limit; light leakage is significantly higher for a partially resolved stellar disk than it would be for an on-axis point source. A shaped pupil could be used instead, but the IWA would be greater than $\sim$3 $\lambda$/D, restricting the target distance at which observations of close-in orbits could be made.

How might our method expand the parameter space of detected exoplanet properties? We have considered binaries that dim by at least 2.5 mag during eclipse. As a result, at a given planet-star contrast and host intrinsic stellar luminosity, our approach allows direct imaging detection of a planet at least 10$\times$ fainter than one around a single host star or non-eclipsing binary system, at the same distance. Physically, for RL planets, 10$\times$ fainter could imply 0.1$\times$ planet albedo or 0.33$\times$ planet size or 3$\times$ planet-star separation (see Eq.~\ref{eq:Rlight}).

Does our method allow access to different types of host stars or binaries than previously explored in exoplanet systems? The Catalogue of Exoplanets in Binary Systems \citep{2016MNRAS.460.3598S} lists detailed properties for 97 binaries with exoplanets, at least five of which were directly imaged. Compared to these, our best target binaries for SL planets are more massive and have earlier-type primaries (i.e., B-type). This difference arises because we selected on higher total binary mass so that any SL planets would be young and therefore bright enough to be detectable.

In general, compared to single star and non-eclipsing binary systems at the same distance, planet luminosity, and planet-star contrast, our SL and RL target binaries would have
stellar luminosities at least 10$\times$ brighter and thus stellar masses  $\sim$2-3$\times$ greater, assuming the canonical main sequence mass-luminosity relation. Therefore, our approach could expand the boundaries of the host stellar mass parameter space. If more exoplanets are discovered around such massive primaries, there are implications for how such planets form and evolve around binaries \citep{2011ApJ...736...89J,2019A&A...626A..99J}.
\section{Conclusions}\label{sec:conclusion}
We investigate the plausibility of a new approach for directly imaging exoplanets. Our idea is to use the eclipse event in eclipsing binary systems as a tool to boost the planet to star flux contrast, i.e., to exploit a natural starshade. During the eclipse, the binary is reduced to a point-like source, making coronagraphic observations possible.

We select 289 binaries where the depth of primary minimum Dmag is $>2.5$ mag, which boosts the planet-star contrast by more than a dex. Thus, at a given observed planet-star contrast and host intrinsic stellar luminosity, we can detect a planet $> 10\times$ fainter in an eclipsing binary, during eclipse, than in other star systems at the same distance. Likewise, we can detect planets of a given intrinsic luminosity around systems whose intrinsic stellar luminosity is $> 10\times$ brighter and whose stellar mass is $\sim2$-$3\times$ greater. In other words, we could directly image exoplanets in a massive binary system at the same contrast level as in a lower mass one.

We consider using this method to detect self-luminous (SL) and reflected light (RL) planets around our binaries.

For the SL planet case, we determine whether 0.5 to 10 $M_J$ planets could be detected during eclipse with current or future coronagraphs.  SL planets are easiest to detect in young systems, given that the thermal emission of the planet decreases with time. Therefore, we select on the age of the binary, as well as the infrared brightness of the planet at the distance of the binary and the planet-star contrast. Because we lack measured ages for our best target binaries, which are classical Algols, we use Algol models \citep{2010AIPC.1314...45V,2017A&A...599A..84M} to constrain a plausible age interval from the binary total mass.

Using these criteria, we identify two targets: U Cep and AC Sct. Around them, we might detect $\sim$4.5$M_J$ and $\sim$9$M_J$ SL planets with current ground- or near-future space-based instruments, respectively. With future ground-based facilities, these limits reduce to 3-4$M_J$ and 6-8$M_J$. Because of our Dmag $> 2.5$ and age criteria, these targets possess larger total masses ($>4.2M_J$) and earlier-type (B7-, B9-) primaries than typical of known host binaries. As noted above, our method puts such massive stellar systems within the reach of coronagraphic observations; targeting these systems would expand the host stellar parameter space for testing SL planet formation and evolution.

For RL planets, the advantage of using eclipsing binaries is that both binary stars continue to illuminate the planet while the planet-star contrast is increased during the eclipse. To find the best targets in this case, we focus on only the nearest (within $\sim$50 pc) eclipsing binaries in our sample: RR Cae, V1412 Aql, and RT Pic, for which the contrast boost during eclipse is 1.32, 1.05, and 1.04 dex, respectively. We assume that a large (30-meter) ground-based telescope and intermediate (2- to 8-meter) space-based telescopes will be available in the future.

We consider a Jupiter-like, Venus-like, and habitable Earth-like planet, estimating the change in detection limit with exposure time in the $J$-band with future ground-based telescope and in the $V$-band with space telescopes. For a ``Jupiter'' at 20 mas in all three target binaries, a detection is achieved in less than $\sim$10 hours with the ground- and space-based telescopes, whereas for a ``Venus'' at 20 mas, detection is possible in the $J$-band around RR Cae and RT Pic and in the $V$-band around RT Pic. Thus, directly imaging these Jupiter- and Venus-like planets is within the capabilities of planned facilities. 

Detection of a habitable Earth-like planet remains a challenge. In a less than $\sim$10 hour exposure in the $J$-band with the ground-based telescope or in the $V$-band with a 6- to 8-meter space telescope during the eclipse, this ``Earth'' would be bright enough to detect if it orbited RT Pic. The planet-star contrast of $1.5\times10^{-9}$ would be achievable from space. The planet-star separation of 13 mas is equivalent to 0.56 $\lambda$/D and 0.74 $\lambda$/D for the future 6-m and 8-m space-based telescopes, respectively. Given that current coronagraphs deliver deep contrast levels at $\gtrsim$2 $\lambda$/D \citep{2006ApJS..167...81G}, larger apertures and/or new coronagraph advances will be required for detection.

\acknowledgments

This work has made use of data from the European Space Agency (ESA) mission {\it Gaia} (\url{https://www.cosmos.esa.int/gaia}), processed by the {\it Gaia} Data Processing and Analysis Consortium (DPAC, \url{https://www.cosmos.esa.int/web/gaia/dpac/consortium}). Funding for the DPAC has been provided by national institutions, in particular the institutions participating in the {\it Gaia} Multilateral Agreement. This research has made use of the SIMBAD database, operated at CDS, Strasbourg, France.

We thank the anonymous referee for detailed comments that substantially improved this paper. We are deeply indebted to Maxwell Moe for sharing his expertise on eclipsing binary evolution, the possibility of tertiaries, and how to estimate the ages of Algols, Glenn Schneider for important discussions on the feasibility of this project, and Elizabeth Green, Andrew Odell, and Joel Eaton for their knowledge about eclipsing binary catalogs. Many thanks to Eric Mamajek, Andrej Prsa, Guillermo Torres, Edwin Budding, and Oleg Malkov for useful discussions on the discrepancies between the values of primary eclipse depth, and to Jack Lissauer for providing insight into the stability of planetary orbits around binary systems. AIZ thanks the Dark Cosmology Centre at the Niels Bohr Institute, University of Copenhagen, for its hospitality and support, which greatly furthered the development of this project.
\appendix
\section{Full Sample of Eclisping Binary Systems}\label{sec:appendix}
\begin{longrotatetable}
\begin{deluxetable}{l|rrrrrrlrrll}
\tablecaption{\label{tab:EBlist}}
\tablewidth{500pt}
\tabletypesize{\scriptsize}
\tablehead{
\colhead{Name} & \colhead{RA} & \colhead{Dec} & \colhead{$m$} & \colhead{Dmag} & \colhead{Filter} & \colhead{Ref} & \colhead{Flag} & \colhead{$\pi$} & \colhead{L$_{\text{tot}}$} & Spec Type & Evol Class\\
\colhead{} & \colhead{[hh mm ss]} & \colhead{[dd mm ss]} & \colhead{} & \colhead{} & \colhead{} & \colhead{} & \colhead{} & \colhead{[mas]} & \colhead{[$L_{\bigodot}$]} & \colhead{} &
}
\startdata
LQ Cas & 00 04 10.9597 & +61 42 07.875 & 14.10 & 3.10 & p & M & 0 & 0.62 & 5.21 & \nodata & \nodata \\
V0411 Cas & 00 30 11.2537 & +56 07 47.734 & 13.90 & 2.80 & B & M & 0 & 0.50 & 11.64 & \nodata & \nodata \\
CV Cas & 00 31 54.4680 & +71 41 38.638 & 13.90 & 3.60 & p & M & 0 & 0.86 & 8.23 & \nodata & \nodata \\
GW And & 00 35 09.4069 & +41 40 03.465 & 14.60 & 2.80 & B & M & 0 & 0.28 & \nodata & \nodata & \nodata \\
KQ Cas & 00 38 37.6964 & +58 32 42.210 & 14.10 & 2.90 & p & M & 0 & 0.31 & 20.36 & \nodata & \nodata \\
UU And & 00 43 45.0803 & +30 56 19.664 & 11.20 & 3.00 & V & B & 0 & 1.06 & 10.35 & F5+[K7IV] & SA \\
Y Hyi & 00 45 50.6974 & -78 49 16.812 & 10.40 & 3.60 & VT & B & 1 & 2.35 & 15.38 & A6V+[K3IV] & SA \\
V Tuc & 00 51 46.5986 & -71 59 52.927 & 10.60 & 8.20 & V & B & 1 & 1.86 & 13.35 & B9/A2IV+[G5IV] & SA \\
V0415 Cas & 00 54 31.0203 & +59 24 00.027 & 14.00 & 2.80 & b & M & 0 & 0.40 & 24.20 & \nodata & \nodata \\
V0386 Cas & 00 59 11.2667 & +55 57 20.037 & 14.00 & 2.50 & b & M & 0 & 0.34 & 35.09 & \nodata & \nodata \\
U Cep & 01 02 18.4416 & +81 52 32.080 & 6.86 & 2.54 & VH & B & 0 & 5.04 & \nodata & B7Ve+B8III-IV & SA \\
FK And & 01 07 03.1786 & +37 29 04.825 & 14.00 & 2.80 & p & M & 0 & 0.30 & \nodata & \nodata & \nodata \\
OY Per & 01 34 53.7662 & +53 39 31.606 & 15.00 & 2.50 & p & M & 0 & 0.33 & 9.28 & \nodata & \nodata \\
BL Hyi & 01 41 00.3995 & -67 53 27.469 & 14.90 & 3.10 & V & M & 0 & 7.65 & \nodata & WD+M3-V & S2C \\
V0367 Cas & 01 41 11.2580 & +61 15 35.226 & 14.20 & 3.10 & b & M & 0 & 0.48 & 12.53 & \nodata & \nodata \\
GH Cas & 01 49 03.3762 & +56 16 26.817 & 12.30 & 3.50 & p & M & 0 & 0.87 & 16.72 & F2 & \nodata \\
HS Per & 01 52 11.6089 & +57 06 42.715 & 13.00 & 2.60 & V & M & 0 & 0.89 & 12.51 & A0II-III & \nodata \\
XZ And & 01 56 51.5242 & +42 06 02.181 & 10.15 & 2.54 & V & M & 1 & 1.84 & 21.97 & A1V+G5IV & SA \\
X Tri & 02 00 33.7370 & +27 53 19.205 & 8.55 & 2.72 & V & B & 1 & 4.82 & \nodata & A5V+[G2IV] & SA \\
LL Per & 02 14 51.5373 & +57 29 34.456 & 17.00 & 2.50 & p & M & 0 & 0.22 & \nodata & \nodata & \nodata \\
MO And & 02 23 53.6676 & +39 59 02.767 & 13.20 & 3.00 & p & M & 0 & 0.52 & 15.65 & \nodata & \nodata \\
LM Per & 02 23 56.6594 & +56 17 19.310 & 16.00 & 2.50 & p & M & 0 & 0.21 & \nodata & \nodata & \nodata \\
RW Tri & 02 25 36.1555 & +28 05 50.893 & 12.50 & 2.42 & V & M & 1 & 3.17 & \nodata & WD+M0V & S2C \\
WW Hor & 02 36 11.4500 & -52 19 13.500 & 18.40 & 4.60 & B & M & 0 & 3.07 & \nodata & WD+M6V & S2C \\
Z Per & 02 40 03.2329 & +42 11 57.694 & 9.70 & 2.70 & p & M & 0 & 1.40 & 42.94 & A0V+G0IV-V & SA \\
V For & 02 40 51.4000 & -25 52 14.000 & 14.05 & 2.95 & B & M & 0 & 0.36 & \nodata & \nodata & \nodata \\
V0366 Per & 02 44 10.6082 & +36 35 11.856 & 13.70 & 2.60 & b & M & 0 & 0.89 & 5.50 & \nodata & \nodata \\
V0479 Per & 02 46 05.6524 & +43 17 07.679 & 14.40 & 3.50 & b & M & 0 & 0.12 & \nodata & \nodata & \nodata \\
SS Cet & 02 48 36.2797 & +01 48 26.894 & 9.40 & 3.60 & V & B & 1 & 1.85 & 25.4 & A0+[K0IV] & SA \\
LS Per & 02 57 08.7937 & +37 53 40.477 & 11.50 & 3.10 & p & M & 0 & 0.88 & 30.23 & A7IV & \nodata \\
QU Per & 03 05 54.6745 & +40 41 12.625 & 12.20 & 2.60 & p & M & 0 & 0.54 & 33.60 & A2Ve & \nodata \\
EF Eri & 03 14 13.2527 & -22 35 42.917 & 14.50 & 2.80 & B & M & 0 & 6.27 & \nodata & WD+M9V & S2C \\
UZ For & 03 35 28.6515 & -25 44 21.765 & 18.20 & 2.70 & V & M & 0 & 4.17 & \nodata & WD+M4.5V & S2C \\
WY Per & 03 38 24.1599 & +42 40 39.050 & 11.50 & 2.70 & V & B & 0 & 0.83 & 21.22 & A0+[K25IV] & SA \\
GH Cam & 03 41 31.4326 & +67 04 43.571 & 13.80 & 3.20 & B & M & 0 & 0.43 & 15.70 & \nodata & \nodata \\
HU Per & 03 42 33.1687 & +39 06 03.625 & 10.90 & 3.00 & p & M & 0 & 0.54 & 66.28 & A & \nodata \\
RW Tau & 04 03 54.3166 & +28 07 33.498 & 8.00 & 3.80 & VH & B & 1 & 3.27 & \nodata & B8Ve+K0IV & SA \\
RV Per & 04 10 37.9975 & +34 15 54.883 & 10.23 & 3.75 & V & B & 1 & 0.94 & 37.71 & A2+[G7IV] & SA \\
V0407 Tau & 04 10 58.4464 & +26 18 21.620 & 13.50 & 3.00 & b & M & 0 & 0.40 & \nodata & \nodata & \nodata \\
BN Tau & 04 15 44.8807 & +30 40 08.025 & 10.17 & 4.21 & VT & B & 1 & 0.73 & 16.43 & A7+[K2IV] & S \\
RR Cae & 04 21 05.5631 & -48 39 07.062 & 14.88 & 3.30 & B & M & 0 & 47.16 & \nodata & WD+M5-6V & DW \\
TZ Eri & 04 21 40.3306 & -06 01 09.201 & 9.80 & 2.80 & V & B & 0 & 3.30 & 10.41 & F3+[K5IV] & SA \\
AO Eri & 04 32 00.9396 & -17 44 47.651 & 10.30 & 3.27 & VT & B & 1 & 1.40 & 19.17 & A8+[G8IV] & S \\
AC Tau & 04 37 06.3566 & +01 41 31.158 & 10.30 & 2.80 & V & B & 0 & 1.77 & 10.44 & F0+[K6IV] & SA \\
AQ Cam & 04 51 17.8661 & +54 55 59.950 & 12.00 & 4.00 & p & M & 0 & 0.23 & \nodata & B7 & \nodata \\
ET Ori & 04 55 34.3335 & +01 22 49.564 & 10.13 & 3.14 & VT & B & 1 & 2.75 & 5.87 & G3+[K3IV] & SA \\
EQ Ori & 04 57 14.4652 & -03 36 04.657 & 10.20 & 3.10 & P & B & 0 & 1.65 & \nodata & A0+[G2IV] & SA \\
V0723 Tau & 05 06 46.1891 & +28 35 23.410 & 14.50 & 2.60 & b & M & 0 & 0.60 & 4.30 & \nodata & \nodata \\
FL Ori & 05 07 46.6517 & -02 44 38.247 & 11.40 & 3.20 & P & B & 0 & 1.98 & 5.91 & A3V+[K0IV] & SA \\
XZ Cam & 05 17 12.6742 & +75 50 05.354 & 11.40 & 3.00 & VT & B & 0 & 0.81 & 42.44 & A0+[K0.5 & SA \\
V0666 Ori & 05 54 06.5077 & +17 56 18.152 & 14.00 & 3.00 & b & M & 0 & 0.75 & 9.84 & \nodata & \nodata \\
RT Pic & 06 00 41.2987 & -44 53 50.106 & 9.90 & 2.60 & p & M & 0 & 18.39 & 0.62 & G8V & \nodata \\
LP Gem & 06 05 05.1282 & +26 40 53.386 & 12.50 & 2.50 & p & M & 0 & 0.53 & 35.88 & \nodata & \nodata \\
DX Ori & 06 05 44.7518 & +08 20 55.289 & 13.50 & 2.50 & p & M & 0 & 0.41 & 17.11 & \nodata & \nodata \\
CG Pup & 06 07 02.9478 & -46 44 47.717 & 10.20 & 2.80 & p & M & 0 & 2.05 & 5.81 & \nodata & \nodata \\
V0347 Pup & 06 10 33.6585 & -48 44 25.363 & 13.55 & 2.40 & B & M & 0 & 3.38 & 0.28 & \nodata & \nodata \\
RU CMa & 06 22 33.5874 & -22 41 29.151 & 11.30 & 2.50 & V & M & 1 & 1.36 & \nodata & \nodata & \nodata \\
BO Gem & 06 25 01.2975 & +17 58 12.777 & 11.30 & 3.80 & P & B & 1 & 1.12 & 21.10 & A2+[K3IV] & SA \\
NS CMa & 06 29 09.5867 & -31 15 33.800 & 14.00 & 2.71 & V & M & 0 & 0.40 & 10.97 & \nodata & \nodata \\
V0639 Mon & 06 37 20.1591 & +11 36 31.485 & 16.00 & 2.50 & b & M & 0 & 0.31 & \nodata & \nodata & \nodata \\
MP Gem & 06 48 33.4000 & +19 37 15.000 & 15.40 & 2.60 & p & M & 0 & 0.41 & \nodata & \nodata & \nodata \\
EF Gem & 06 51 01.8159 & +17 29 54.987 & 12.60 & 2.80 & p & M & 0 & 0.41 & 36.42 & \nodata & \nodata \\
CH Mon & 06 51 38.1214 & +05 56 12.268 & 13.10 & 2.80 & p & M & 0 & 0.57 & 17.73 & \nodata & \nodata \\
EH Mon & 06 52 08.7658 & -07 03 52.971 & 13.80 & 2.50 & p & M & 0 & 0.19 & \nodata & \nodata & \nodata \\
RV Lyn & 06 56 11.4299 & +50 51 45.509 & 13.00 & 2.60 & p & M & 0 & 0.66 & 12.38 & \nodata & \nodata \\
BP Mon & 06 56 55.4325 & +05 01 43.820 & 13.10 & 2.60 & p & M & 0 & 0.74 & 11.06 & \nodata & \nodata \\
FS Gem & 06 57 21.2259 & +16 30 13.681 & 13.70 & 2.90 & p & M & 0 & 0.25 & \nodata & \nodata & \nodata \\
UU CMa & 06 58 41.4546 & -18 48 48.187 & 10.00 & 2.50 & p & M & 0 & 1.79 & 9.93 & \nodata & \nodata \\
AC CMa & 07 08 23.2400 & -19 40 01.547 & 12.80 & 2.50 & p & M & 0 & 0.75 & 17.91 & \nodata & \nodata \\
AB CMi & 07 07 57.3118 & +11 58 18.883 & 13.40 & 2.50 & p & M & 0 & 0.26 & \nodata & \nodata & \nodata \\
HO Mon & 07 10 19.2100 & +00 25 28.949 & 11.40 & 2.80 & P & B & 0 & 1.05 & 22.02 & A5+[K2.5IV] & SA \\
AQ Mon & 07 14 17.6748 & -07 13 45.060 & 9.79 & 2.71 & VT & B & 1 & 1.48 & 21.39 & A0+[A4.5] & SA \\
FP Mon & 07 15 08.8358 & -09 57 47.639 & 13.30 & 3.00 & p & M & 0 & 0.85 & 7.05 & F6V & \nodata \\
AF CMa & 07 15 22.7708 & -23 40 42.624 & 12.30 & 3.20 & p & M & 0 & 0.67 & 33.76 & \nodata & \nodata \\
HS Cam & 07 19 14.5151 & +65 57 44.305 & 19.40 & 4.00 & B & M & 0 & 2.19 & \nodata & WD+M4-6V & S2C \\
RY CMi & 07 22 58.1591 & +06 46 35.019 & 11.90 & 3.00 & p & M & 0 & 0.55 & \nodata & \nodata & S \\
CX Pup & 07 41 53.5171 & -22 53 36.924 & 12.20 & 3.00 & p & M & 0 & 0.84 & \nodata & \nodata & \nodata \\
V0681 Mon & 07 52 21.9573 & -01 17 25.149 & 13.00 & 2.50 & b & M & 0 & 0.81 & 16.18 & \nodata & \nodata \\
TU Mon & 07 53 19.7487 & -03 02 31.137 & 9.00 & 2.70 & VH & B & 1 & 0.61 & \nodata & B5V+A5III: & SH \\
MT Pup & 07 54 11.8928 & -14 39 16.633 & 15.00 & 4.00 & p & M & 0 & 1.01 & \nodata & em & \nodata \\
FG Pup & 08 04 31.6979 & -24 02 53.406 & 13.00 & 3.00 & p & M & 0 & 0.46 & 22.14 & \nodata & \nodata \\
IO Pup & 08 06 57.9454 & -25 49 08.288 & 13.50 & 2.50 & p & M & 0 & 0.48 & \nodata & \nodata & \nodata \\
AV Vel & 08 07 01.3411 & -47 38 46.007 & 11.90 & 2.70 & p & M & 0 & 0.89 & 14.32 & \nodata & \nodata \\
FI Pup & 08 08 14.7308 & -20 17 14.882 & 14.00 & 2.50 & p & M & 0 & 0.24 & \nodata & \nodata & \nodata \\
XZ Pup & 08 13 31.0553 & -23 57 11.387 & 7.75 & 2.51 & V & M & 1 & 2.54 & \nodata & A0 & SA \\
FQ Pup & 08 18 10.5357 & -23 48 40.571 & 14.00 & 2.50 & p & M & 0 & 0.43 & 8.11 & \nodata & \nodata \\
FR Pup & 08 21 19.9553 & -22 18 45.703 & 13.50 & 3.00 & p & M & 0 & 0.27 & \nodata & \nodata & \nodata \\
DR Lyn & 08 24 24.4675 & +50 00 50.891 & 11.60 & 2.70 & V & M & 0 & 1.24 & 6.42 & \nodata & \nodata \\
DE Hya & 08 27 47.5470 & +05 38 58.939 & 11.00 & 3.00 & V & B & 0 & \nodata & \nodata & A2+[K2IV] & SA \\
SY Hya & 08 29 51.6786 & -09 23 57.496 & 10.70 & 2.90 & VT & B & 0 & 0.93 & \nodata & A5+[K2IV] & SA \\
VW Hya & 08 33 51.0321 & -14 39 53.891 & 10.50 & 3.60 & V & B & 1 & 0.80 & \nodata & A3+[G4IV] & SA \\
RY Cnc & 08 39 54.6211 & +19 49 18.864 & 12.99 & 2.53 & V & B & 0 & 0.85 & 7.46 & B8+[G5IV] & S \\
TY Cnc & 08 47 09.6962 & +08 24 24.582 & 12.70 & 3.00 & p & M & 0 & 0.69 & \nodata & \nodata & \nodata \\
TU Cnc & 08 52 16.6529 & +09 05 18.781 & 9.90 & 2.50 & P & B & 0 & 1.42 & 28.54 & A0+[G8 & SA \\
AC UMa & 08 55 54.1228 & +64 58 14.999 & 10.30 & 3.70 & V & B & 0 & 1.24 & 41.14 & A2+[K6IV] & SA \\
RX Hya & 09 05 41.1657 & -08 15 39.727 & 8.90 & 2.70 & V & B & 0 & 3.71 & 8.98 & A8+[K5IV] & SA \\
TY Hya & 09 29 02.3991 & +05 34 28.269 & 10.50 & 3.00 & P & B & 0 & 0.76 & 37.79 & A0+[K1IV] & SA \\
Y Leo & 09 36 51.8058 & +26 13 57.643 & 10.09 & 3.11 & V & M & 1 & 2.49 & 11.94 & A5V+K5V & SA \\
DX Vel & 09 51 45.9871 & -50 53 23.716 & 10.08 & 3.42 & VT & B & 1 & 1.85 & 12.12 & A5+[G1IV] & SA \\
IV Car & 10 02 36.6688 & -58 57 22.460 & 11.30 & 2.80 & p & M & 0 & 1.29 & \nodata & \nodata & \nodata \\
HL Car & 10 04 42.7497 & -61 50 35.084 & 11.00 & 3.00 & p & M & 0 & 0.77 & 53.09 & \nodata & \nodata \\
TT Vel & 10 20 16.4606 & -46 14 04.436 & 10.10 & 3.24 & VT & B & 1 & 1.95 & 12.54 & A5+[F8.5] & SA \\
AW Car & 10 32 35.0208 & -60 46 10.653 & 12.20 & 4.30 & p & M & 0 & 0.34 & 39.11 & A3 & \nodata \\
FV Car & 10 42 31.9233 & -61 57 55.072 & 12.90 & 2.90 & p & M & 0 & 0.57 & 60.55 & \nodata & S \\
PT Car & 11 03 17.6000 & -74 56 37.000 & 12.80 & 3.20 & p & M & 0 & \nodata & \nodata & \nodata & \nodata \\
EH Car & 11 04 23.8163 & -61 18 23.020 & 12.50 & 2.50 & p & M & 0 & 0.30 & 47.39 & \nodata & \nodata \\
AN UMa & 11 04 25.6556 & +45 03 13.941 & 14.90 & 5.30 & B & M & 0 & 3.10 & \nodata & \nodata & S2C \\
DE Car & 11 06 13.6515 & -60 47 32.672 & 11.20 & 2.60 & p & M & 1 & 0.94 & 17.85 & A & SA \\
V0443 Cen & 11 25 22.5000 & -59 37 40.000 & 12.80 & 2.50 & b & M & 0 & \nodata & \nodata & \nodata & \nodata \\
AB Cen & 11 26 21.2641 & -58 24 58.136 & 10.60 & 2.60 & p & M & 0 & 0.62 & 85.93 & G0 & \nodata \\
V0646 Cen & 11 36 58.7768 & -53 12 35.414 & 9.00 & 2.70 & b & M & 0 & 1.40 & \nodata & B8IV & S \\
V0348 Cen & 11 48 28.9566 & -43 46 53.264 & 10.70 & 2.70 & b & M & 0 & 1.26 & 14.73 & A & \nodata \\
Z Crv & 12 29 41.7139 & -23 38 03.878 & 14.00 & 2.50 & V & M & 0 & 1.21 & 1.32 & \nodata & \nodata \\
RR Crv & 12 30 26.9456 & -17 58 53.449 & 11.10 & 2.80 & p & M & 0 & 1.33 & 12.15 & \nodata & \nodata \\
BP Mus & 12 50 37.7277 & -71 46 18.698 & 9.60 & 3.00 & V & M & 0 & 1.75 & 21.65 & A0,5/1,5V+G5III & \nodata \\
CT Cen & 13 10 43.0552 & -58 16 38.629 & 10.30 & 2.70 & p & M & 0 & 1.56 & 41.87 & A3(ea) & SA \\
CY Cen & 13 12 13.0891 & -52 52 51.852 & 13.30 & 2.80 & p & M & 0 & 0.70 & 6.89 & \nodata & \nodata \\
UW Vir & 13 15 20.7356 & -17 28 16.924 & 8.84 & 3.16 & V & B & 0 & 4.27 & 10.35 & A4+[K3IV] & SA \\
BM Cen & 13 17 38.4543 & -56 17 09.765 & 13.20 & 2.80 & p & M & 0 & 0.69 & 7.51 & \nodata & \nodata \\
BP Cen & 13 19 08.6963 & -49 55 04.751 & 12.20 & 3.80 & p & M & 0 & 1.14 & \nodata & \nodata & \nodata \\
SX Hya & 13 44 37.3594 & -26 46 48.376 & 8.60 & 4.00 & VH & B & 1 & 3.86 & 16.23 & A3+[G9IV] & SA \\
IQ Cen & 14 00 34.4391 & -54 16 15.669 & 12.70 & 3.00 & p & M & 0 & 0.51 & 28.32 & \nodata & \nodata \\
AN Cir & 14 04 21.9000 & -65 38 05.000 & 13.80 & 2.80 & p & M & 0 & \nodata & \nodata & \nodata & \nodata \\
GK Vir & 14 15 36.4126 & +01 17 18.227 & 17.01 & 5.99 & V & M & 0 & 2.11 & \nodata & DA0+M3-5V & DW \\
TU Lup & 14 43 40.6725 & -49 50 24.752 & 12.40 & 3.80 & p & M & 0 & 0.72 & 11.73 & \nodata & \nodata \\
W Cir & 15 11 47.0000 & -55 38 12.000 & 13.80 & 2.70 & b & M & 0 & \nodata & \nodata & \nodata & \nodata \\
EV Lup & 15 16 08.0026 & -44 19 20.380 & 9.80 & 2.70 & p & M & 0 & 1.05 & 33.65 & \nodata & \nodata \\
TW Dra & 15 33 51.0598 & +63 54 25.706 & 8.00 & 2.50 & p & M & 1 & 6.01 & 22.75 & A8V+K0III & SA \\
LX Ser & 15 38 00.0940 & +18 52 03.246 & 13.30 & 4.10 & B & M & 0 & 2.03 & \nodata & WD+M32V & S2C \\
HH Nor & 15 43 30.1810 & -51 50 47.542 & 9.70 & 3.05 & VT & B & 1 & 2.22 & 15.31 & A3+[G8IV] & S \\
YY Nor & 15 47 58.4618 & -57 24 42.626 & 13.20 & 2.90 & p & M & 0 & 0.76 & 12.13 & \nodata & \nodata \\
HK Nor & 15 58 54.7069 & -51 41 54.960 & 12.60 & 2.60 & p & M & 0 & 0.88 & 3.69 & \nodata & \nodata \\
HS Nor & 16 01 44.9307 & -57 27 33.167 & 12.60 & 2.60 & p & M & 0 & 0.41 & 25.41 & \nodata & \nodata \\
BH Aps & 16 08 40.2241 & -76 50 28.071 & 10.80 & 2.60 & p & M & 0 & 1.00 & \nodata & \nodata & \nodata \\
CE Nor & 16 11 13.2040 & -60 02 20.928 & 11.50 & 3.00 & p & M & 0 & 0.60 & \nodata & \nodata & \nodata \\
CC Her & 16 17 38.8937 & +08 56 02.610 & 10.20 & 2.90 & p & M & 0 & 2.82 & 10.18 & A1V(pSr) & SA \\
UU Nor & 16 20 21.5635 & -53 44 52.236 & 11.60 & 2.70 & p & M & 0 & 0.37 & 78.91 & \nodata & \nodata \\
V0704 Ara & 16 56 16.9143 & -55 27 58.361 & 12.90 & 2.70 & b & M & 0 & 0.46 & 13.09 & \nodata & \nodata \\
UU Oph & 16 57 22.6406 & -25 47 58.563 & 10.00 & 2.50 & VH & B & 0 & 0.76 & \nodata & A0+[K2IV] & SA \\
V0588 Sco & 16 57 49.7872 & -40 29 59.953 & 14.00 & 2.50 & b & M & 0 & 0.58 & 7.09 & \nodata & \nodata \\
V0416 Ara & 16 58 06.6680 & -55 30 12.252 & 12.60 & 2.90 & b & M & 0 & 0.36 & 28.39 & \nodata & \nodata \\
IM Sco & 17 00 13.2531 & -31 27 57.609 & 13.30 & 3.20 & p & M & 0 & 0.64 & 4.29 & G2V & \nodata \\
V1194 Oph & 17 00 21.1176 & -21 51 22.343 & 17.60 & 2.70 & p & M & 0 & 0.24 & \nodata & \nodata & \nodata \\
V735 Oph & 17 07 48.9680 & +09 33 09.206 & 10.30 & 2.96 & VT & B & 0 & 1.78 & 11.25 & A0+[G1IV] & SA \\
V0460 Sco & 17 08 50.4000 & -32 54 05.000 & 13.40 & 2.60 & b & M & 0 & \nodata & \nodata & \nodata & \nodata \\
V1453 Oph & 17 08 52.7155 & -19 01 08.933 & 17.40 & 3.00 & p & M & 0 & 0.30 & \nodata & \nodata & \nodata \\
FT Sco & 17 10 55.2000 & -32 41 09.000 & 13.80 & 2.70 & p & M & 0 & \nodata & \nodata & \nodata & \nodata \\
V1578 Oph & 17 12 14.2156 & -20 14 54.864 & 16.40 & 3.40 & p & M & 0 & 0.11 & \nodata & \nodata & \nodata \\
V1590 Oph & 17 12 22.2904 & -16 24 34.019 & 16.00 & 2.90 & p & M & 0 & 0.26 & \nodata & \nodata & \nodata \\
TU Her & 17 13 35.3658 & +30 42 36.042 & 10.88 & 2.82 & V & M & 1 & 1.96 & 9.10 & F0III/IV & SA \\
V0621 Sco & 17 14 36.8000 & -41 04 15.000 & 12.30 & 2.70 & b & M & 0 & \nodata & \nodata & \nodata & \nodata \\
V0441 Oph & 17 20 52.7094 & -17 20 05.261 & 11.60 & 3.00 & b & M & 0 & 1.26 & 16.83 & A0 & \nodata \\
LT Ara & 17 22 30.1256 & -47 04 02.989 & 10.80 & 4.60 & p & M & 0 & 0.87 & 39.25 & A & \nodata \\
V0506 Sco & 17 23 12.4000 & -39 11 40.000 & 13.30 & 2.70 & b & M & 0 & 1.02 & \nodata & \nodata & \nodata \\
V0510 Sco & 17 27 37.2557 & -42 58 53.276 & 13.10 & 3.00 & b & M & 0 & 0.42 & 35.92 & \nodata & \nodata \\
V0511 Sco & 17 28 22.5406 & -42 02 39.060 & 12.50 & 3.50 & b & M & 0 & 0.79 & 18.64 & \nodata & \nodata \\
V0532 Oph & 17 32 42.6073 & -21 51 40.760 & 12.60 & 2.90 & b & M & 0 & 0.17 & \nodata & \nodata & \nodata \\
RW Ara & 17 34 49.2233 & -57 08 50.566 & 8.85 & 2.60 & V & M & 1 & 2.53 & \nodata & A1IV & SA \\
V0525 Sco & 17 39 57.7671 & -40 29 13.385 & 13.80 & 2.60 & b & M & 0 & 0.58 & 11.51 & \nodata & \nodata \\
V0529 Oph & 17 41 56.6000 & -28 18 10.000 & 13.20 & 3.30 & b & M & 0 & \nodata & \nodata & \nodata & \nodata \\
AK Ser & 17 42 04.4088 & -13 53 11.971 & 10.80 & 2.80 & p & M & 1 & 1.33 & \nodata & A5 & SA \\
V0755 Sgr & 17 48 46.1804 & -25 53 09.993 & 13.50 & 2.50 & b & M & 0 & 0.61 & \nodata & \nodata & \nodata \\
V0765 Sgr & 17 51 54.0858 & -28 26 32.028 & 12.80 & 3.70 & b & M & 0 & 0.37 & 46.29 & \nodata & \nodata \\
V913 Oph & 17 55 03.5918 & +14 10 39.254 & 11.50 & 3.00 & P & B & 0 & 2.14 & 5.19 & A5+[G5IV] & SA \\
V391 Oph & 17 58 09.1304 & +04 39 27.610 & 11.50 & 3.50 & P & B & 1 & 0.70 & 12.93 & A1+[G5IV] & SA \\
V2301 Oph & 18 00 35.5318 & +08 10 13.920 & 16.00 & 6.00 & V & M & 0 & 8.24 & \nodata & WD+M6V & S2C \\
V0573 Oph & 18 06 05.3800 & +02 05 43.828 & 13.10 & 3.00 & b & M & 0 & 0.48 & 19.51 & \nodata & \nodata \\
V1178 Sgr & 18 11 03.8499 & -30 30 56.884 & 12.80 & 2.50 & p & M & 0 & 0.08 & \nodata & \nodata & \nodata \\
AG Pav & 18 12 25.6276 & -62 33 21.859 & 10.20 & 2.80 & p & M & 0 & 1.62 & 10.48 & \nodata & \nodata \\
V1282 Sgr & 18 13 34.6857 & -34 28 42.944 & 11.90 & 4.30 & p & M & 0 & 1.08 & 13.03 & \nodata & \nodata \\
V1959 Sgr & 18 16 43.0576 & -25 31 20.268 & 12.60 & 3.20 & p & M & 0 & 0.52 & 13.03 & \nodata & \nodata \\
AI Sgr & 18 16 45.0943 & -21 34 48.529 & 11.80 & 3.20 & p & M & 0 & 0.43 & 87.41 & \nodata & \nodata \\
V0710 Sgr & 18 16 46.8091 & -36 25 32.160 & 13.30 & 3.00 & b & M & 0 & 0.51 & \nodata & \nodata & \nodata \\
V3171 Sgr & 18 22 44.2273 & -34 13 32.205 & 16.00 & 3.20 & p & M & 0 & 0.16 & \nodata & \nodata & \nodata \\
V2537 Sgr & 18 23 00.307 & -32 34 03.030 & 12.30 & 2.50 & p & M & 0 & \nodata & \nodata & \nodata & \nodata \\
V1654 Sgr & 18 23 17.1000 & -23 15 10.000 & 13.20 & 2.60 & p & M & 0 & \nodata & \nodata & \nodata & \nodata \\
V0586 Oph & 18 27 13.9427 & 04 17 15.319 & 13.30 & 2.50 & b & M & 0 & 0.77 & 14.19 & \nodata & \nodata \\
V3421 Sgr & 18 29 11.6135 & -32 23 55.711 & 16.50 & 2.60 & p & M & 0 & 0.28 & \nodata & \nodata & \nodata \\
V3476 Sgr & 18 30 49.5100 & -32 57 51.000 & 14.90 & 2.50 & p & M & 0 & \nodata & \nodata & \nodata & \nodata \\
EY Ser & 18 30 49.8137 & +05 30 17.629 & 14.50 & 3.00 & p & M & 0 & 0.54 & 12.60 & \nodata & \nodata \\
V4727 Sgr & 18 33 30.7141 & -28 58 51.239 & 10.60 & 3.20 & V & M & 0 & 1.15 & 14.61 & \nodata & \nodata \\
V3622 Sgr & 18 35 46.2499 & -34 52 30.616 & 14.80 & 3.00 & p & M & 0 & 0.21 & \nodata & \nodata & \nodata \\
V0372 Sct & 18 38 03.6563 & -04 01 06.118 & 14.20 & 2.60 & b & M & 0 & 0.35 & 54.58 & \nodata & \nodata \\
V3696 Sgr & 18 38 46.8768 & -34 39 03.802 & 16.30 & 3.00 & p & M & 0 & 0.23 & \nodata & \nodata & \nodata \\
BO Her & 18 40 30.0965 & +24 55 42.768 & 10.70 & 3.10 & V & M & 1 & 1.29 & 17.08 & A7V & SA \\
V3752 Sgr & 18 40 43.9834 & -35 32 26.340 & 15.90 & 2.60 & p & M & 0 & 0.29 & \nodata & \nodata & \nodata \\
RR Dra & 18 41 47.4012 & +62 40 34.944 & 10.00 & 3.30 & V & B & 0 & 2.17 & 20.23 & A2+[G8IV] & SA \\
V1933 Sgr & 18 45 10.1332 & -19 47 19.960 & 12.50 & 2.50 & p & M & 0 & 0.62 & 22.20 & \nodata & \nodata \\
AC Sct & 18 46 01.4709 & -10 14 55.815 & 10.00 & 2.60 & V & B & 0 & 1.01 & 63.82 & B9+[G0IV] & SA \\
V0795 Aql & 18 47 06.8639 & +11 40 11.052 & 13.40 & 3.00 & b & M & 0 & 0.40 & 20.70 & \nodata & \nodata \\
DH Her & 18 47 34.5584 & +22 50 45.795 & 9.40 & 2.60 & V & M & 1 & 0.33 & \nodata & A5 & SA \\
V0370 Sct & 18 48 31.2299 & -05 42 37.449 & 15.30 & 3.20 & b & M & 0 & 0.46 & 8.35 & \nodata & \nodata \\
AD Her & 18 50 00.3000 & +20 43 16.509 & 9.38 & 10.74 & V & B & 1 & 1.71 & 37.58 & A4V+[G9IV] & SA \\
FN Sct & 18 50 46.7234 & -05 12 07.916 & 12.70 & 2.50 & B & M & 0 & 0.35 & 108.38 & B3Ve & \nodata \\
CT Sct & 18 54 21.6365 & -06 00 16.696 & 10.02 & 4.17 & V & B & 1 & 0.26 & 631.53 & B7Vn+[A3] & SA \\
LP Her & 18 55 41.2099 & +12 15 28.400 & 13.00 & 3.20 & p & M & 0 & 0.65 & 16.56 & \nodata & \nodata \\
EH Sct & 18 57 39.7600 & -06 52 59.700 & 14.80 & 2.70 & p & M & 0 & \nodata & \nodata & \nodata & \nodata \\
BN Sct & 18 58 42.8687 & -08 20 28.364 & 11.40 & 3.20 & p & M & 0 & 0.82 & 44.06 & A0Ib & \nodata \\
HY Aql & 19 02 47.6109 & -06 13 33.067 & 12.80 & 2.80 & p & M & 0 & 0.60 & 17.21 & \nodata & \nodata \\
FK Aql & 19 04 18.5775 & +02 46 47.244 & 11.10 & 2.40 & VT & B & 0 & 1.38 & 19.77 & B9+[G6IV] & SA \\
V1074 Sgr & 19 05 23.6885 & -18 47 06.864 & 12.70 & 2.50 & p & M & 0 & 1.26 & 7.49 & F4III & \nodata \\
V1109 Aql & 19 05 40.1991 & +14 14 43.686 & 17.00 & 2.50 & p & M & 0 & 0.19 & \nodata & \nodata & \nodata \\
EP Dra & 19 07 06.1869 & +69 08 43.874 & 17.60 & 2.90 & V & M & 0 & 2.59 & \nodata & \nodata & S2C \\
V0413 Lyr & 19 07 16.4316 & +30 19 25.590 & 14.50 & 2.60 & b & M & 0 & 0.23 & 21.65 & \nodata & \nodata \\
V0449 Lyr & 19 07 37.0709 & +44 00 19.797 & 12.50 & 3.50 & b & M & 0 & 0.18 & \nodata & \nodata & \nodata \\
V0415 Sgr & 19 07 37.0938 & -23 37 08.543 & 12.30 & 3.00 & b & M & 0 & 0.55 & 38.86 & \nodata & \nodata \\
V1112 Aql & 19 07 39.1500 & -00 23 07.100 & 14.00 & 2.50 & p & M & 0 & \nodata & \nodata & \nodata & \nodata \\
KK Dra & 19 07 56.6643 & +59 23 52.065 & 11.80 & 3.00 & V & M & 0 & 1.69 & 3.63 & \nodata & \nodata \\
V0407 Aql & 19 11 10.8162 & +01 08 52.145 & 13.30 & 2.50 & b & M & 0 & 1.24 & 4.33 & \nodata & \nodata \\
NS Lyr & 19 11 13.9662 & +36 10 35.321 & 14.00 & 2.80 & p & M & 0 & 0.48 & 9.32 & \nodata & \nodata \\
V354 Sgr & 19 13 23.4427 & -18 29 00.014 & 10.70 & 3.00 & VT & B & 1 & 1.50 & 19.05 & F8+[K9IV] & SA \\
V1103 Cyg & 19 14 57.6721 & +46 10 01.622 & 15.00 & 2.50 & p & M & 0 & 0.30 & 10.29 & \nodata & \nodata \\
RV Lyr & 19 16 17.9644 & +32 25 14.969 & 11.50 & 3.10 & P & B & 0 & 1.02 & 22.01 & A5+[K3IV] & SA \\
YZ Aql & 19 16 46.2253 & -00 36 17.152 & 10.50 & 3.70 & V & B & 1 & 0.92 & 45.94 & A3+[K5IV] & SA \\
V342 Aql & 19 17 03.4768 & +09 20 38.546 & 9.50 & 3.40 & P & B & 0 & 3.21 & 25.59 & A4II+[K0IV] & SA \\
U Sge & 19 18 48.4083 & +19 36 37.720 & 6.50 & 2.60 & V & B & 1 & 3.62 & \nodata & B8III+K & SA \\
FR Vul & 19 36 24.8396 & +26 45 56.558 & 9.91 & 2.54 & VT & B & 1 & 1.80 & 16.20 & A2+[G7IV] & SA \\
V0418 Aql & 19 36 39.4426 & +03 57 00.470 & 12.80 & 3.20 & b & M & 0 & 1.06 & 6.72 & \nodata & \nodata \\
LT Aql & 19 38 49.7896 & +06 34 59.039 & 12.40 & 2.50 & p & M & 0 & 0.59 & 30.27 & B9III/IV & \nodata \\
V0932 Cyg & 19 39 05.8610 & +33 29 01.674 & 14.00 & 2.60 & b & M & 0 & 0.22 & 38.84 & \nodata & \nodata \\
EN Cyg & 19 40 09.4635 & +29 16 23.100 & 12.90 & 3.20 & p & M & 0 & 1.06 & 7.76 & \nodata & \nodata \\
V2168 Sgr & 19 42 37.5266 & -38 39 55.648 & 12.50 & 2.50 & p & M & 0 & 0.76 & 7.23 & \nodata & \nodata \\
SY Cyg & 19 46 34.3155 & +32 42 18.463 & 10.70 & 3.50 & P & B & 0 & 0.94 & 39.95 & A3+[K6IV] & SA \\
V1033 Aql & 19 46 44.7712 & +14 21 57.379 & 15.00 & 2.50 & p & M & 0 & 0.20 & \nodata & \nodata & \nodata \\
V0814 Cyg & 19 49 50.7520 & +36 34 27.659 & 15.00 & 3.00 & b & M & 0 & 0.17 & \nodata & \nodata & \nodata \\
V0689 Cyg & 19 51 06.5892 & +36 51 32.721 & 14.00 & 2.50 & b & M & 0 & 0.65 & 6.35 & A5V & \nodata \\
V524 Sgr & 19 53 14.5011 & -14 54 38.882 & 9.96 & 2.54 & VT & B & 1 & 2.16 & 12.25 & F8III+[K3IV] & SA \\
V0340 Aql & 19 55 56.4890 & +15 51 07.167 & 11.50 & 2.60 & b & M & 0 & 0.88 & 16.30 & F & \nodata \\
BO Vul & 19 56 29.0711 & +23 54 45.011 & 10.50 & 2.80 & p & M & 0 & 2.79 & 9.76 & F0+G0IV & SA \\
PV Cyg & 19 56 29.1748 & +37 43 09.124 & 12.70 & 2.80 & p & M & 0 & 0.58 & 13.71 & A1 & \nodata \\
V0691 Cyg & 19 57 18.4468 & +40 02 55.712 & 15.50 & 2.50 & b & M & 0 & 0.35 & 6.50 & \nodata & \nodata \\
QT Cyg & 19 58 07.3701 & +38 49 27.887 & 14.80 & 2.60 & p & M & 0 & 0.26 & 15.77 & A5 & \nodata \\
V4140 Sgr & 19 58 49.7022 & -38 56 13.225 & 15.50 & 2.50 & p & M & 0 & 1.67 & \nodata & WD+M6-7 & S2C \\
QX Sge & 19 59 36.7480 & +20 48 14.599 & 20.40 & 2.60 & V & M & 0 & \nodata & \nodata & \nodata & \nodata \\
V0698 Cyg & 19 59 53.3495 & +36 16 39.998 & 12.20 & 2.70 & b & M & 0 & 0.28 & \nodata & B2 & \nodata \\
V1174 Cyg & 20 03 55.9791 & +31 15 39.947 & 14.50 & 2.50 & p & M & 0 & 0.31 & 38.67 & \nodata & \nodata \\
WW Cyg & 20 04 02.7111 & +41 35 16.462 & 10.02 & 3.24 & V & B & 0 & 0.80 & \nodata & B7V+[G1IV] & SA \\
EX Vul & 20 04 47.4096 & +22 19 21.861 & 12.49 & 2.61 & B & M & 0 & 0.51 & 52.47 & ea & \nodata \\
SW Cyg & 20 06 57.9310 & +46 17 58.147 & 9.24 & 2.59 & V & M & 1 & 2.12 & 33.46 & A2e+K0 & SA \\
V1037 Cyg & 20 08 49.5433 & +35 14 55.507 & 14.70 & 2.60 & p & M & 0 & 0.51 & 8.05 & \nodata & \nodata \\
QY Aql & 20 09 28.8293 & +15 18 44.718 & 11.40 & 3.20 & VH & B & 0 & 0.78 & 57.18 & F0+/K3IV/ & SA \\
V1412 Aql & 20 13 55.6789 & +06 42 44.826 & 15.67 & 2.63 & V & M & 0 & 43.62 & \nodata & DC7 & \nodata \\
V0445 Aql & 20 19 38.8300 & +06 08 47.400 & 13.00 & 2.50 & b & M & 0 & \nodata & \nodata & \nodata & \nodata \\
V Sge & 20 20 14.6910 & +21 06 17.17 & 8.60 & 5.30 & V & M & 0 & 0.42 & \nodata & WN5 & S2C \\
V1320 Cyg & 20 21 22.0535 & +39 19 57.924 & 15.90 & 2.60 & p & M & 0 & 0.44 & 14.06 & \nodata & \nodata \\
V0445 Cyg & 20 28 18.9582 & +38 17 43.226 & 11.70 & 3.80 & b & M & 0 & 1.28 & 7.36 & \nodata & \nodata \\
V1051 Cyg & 20 31 00.5678 & +56 46 31.159 & 14.40 & 2.80 & p & M & 0 & 0.52 & 5.52 & \nodata & \nodata \\
LL Vul & 20 31 43.3028 & +25 38 36.303 & 16.00 & 3.00 & p & M & 0 & 0.09 & \nodata & \nodata & \nodata \\
HN Del & 20 33 45.1355 & +11 03 40.875 & 14.50 & 2.50 & p & M & 0 & 0.59 & 4.96 & \nodata & \nodata \\
SY Vul & 20 36 21.8687 & +23 51 51.038 & 13.00 & 3.00 & p & M & 0 & 0.28 & 29.26 & \nodata & \nodata \\
KK Del & 20 36 42.7618 & +17 28 48.078 & 15.00 & 2.50 & p & M & 0 & 0.98 & \nodata & \nodata & \nodata \\
W Del & 20 37 40.0857 & +18 17 03.752 & 9.69 & 2.64 & V & M & 1 & 1.23 & \nodata & B9.5Ve+G5IV & SA \\
V1204 Cyg & 20 44 10.7490 & +46 49 03.756 & 14.50 & 3.00 & p & M & 0 & 0.48 & 6.93 & \nodata & \nodata \\
V1843 Cyg & 20 46 27.7000 & +34 03 53.000 & 15.70 & 2.80 & B & M & 0 & 0.20 & \nodata & \nodata & \nodata \\
V0398 Cyg & 20 46 50.7555 & +34 12 03.717 & 12.50 & 2.50 & b & M & 0 & 0.70 & 14.20 & \nodata & \nodata \\
DW Cep & 20 51 39.6899 & +62 48 50.292 & 10.25 & 2.65 & VT & B & 1 & 1.21 & 20.02 & B8+[F5] & SA \\
V1870 Cyg & 20 51 41.0469 & +35 44 08.101 & 14.30 & 2.90 & B & M & 0 & 0.86 & 4.37 & \nodata & \nodata \\
V1884 Cyg & 20 56 07.1165 & +33 39 07.145 & 15.20 & 2.50 & B & M & 0 & 0.25 & 27.25 & \nodata & \nodata \\
V1718 Cyg & 21 02 01.2000 & +41 31 57.000 & 14.70 & 2.80 & p & M & 0 & \nodata & \nodata & \nodata & \nodata \\
V0377 Cyg & 21 03 12.1952 & +29 07 11.878 & 14.40 & 2.60 & b & M & 0 & 0.28 & 33.71 & \nodata & \nodata \\
LY Del & 21 06 26.0978 & +19 24 36.422 & 10.40 & 3.10 & V & M & 0 & 0.54 & 129.27 & \nodata & \nodata \\
V1539 Oph & 21 11 20.0900 & -20 29 04.967 & 15.60 & 3.20 & p & M & 0 & 0.40 & 5.33 & \nodata & \nodata \\
V1960 Cyg & 21 12 41.9861 & +37 32 28.212 & 15.30 & 2.50 & B & M & 0 & 0.26 & 12.01 & \nodata & \nodata \\
V0534 Cyg & 21 21 08.4383 & +44 15 04.382 & 13.70 & 3.00 & b & M & 0 & 0.60 & 8.49 & \nodata & \nodata \\
V1727 Cyg & 21 31 26.2132 & +47 17 24.514 & 15.60 & 2.90 & V & M & 0 & 0.55 & \nodata & pec(e) & S2L \\
U Gru & 21 31 48.7723 & -45 02 42.324 & 11.00 & 4.00 & P & B & 1 & 1.45 & 18.67 & A5+[K0IV] & SA \\
V1667 Cyg & 21 32 33.3186 & +34 27 06.581 & 14.90 & 2.50 & B & M & 0 & 3.28 & 1.50 & \nodata & \nodata \\
BQ Peg & 21 34 20.6566 & +20 57 18.118 & 13.70 & 2.60 & p & M & 0 & 0.61 & 8.70 & \nodata & \nodata \\
V0705 Cyg & 21 34 56.6008 & +43 01 28.616 & 13.60 & 2.80 & b & M & 0 & 0.47 & 18.30 & \nodata & \nodata \\
V1618 Cyg & 21 45 54.1084 & +38 45 26.737 & 15.40 & 2.60 & p & M & 0 & 0.30 & 8.31 & \nodata & \nodata \\
CW Peg & 21 48 27.6008 & +28 06 29.191 & 11.80 & 3.43 & V & M & 1 & 1.02 & \nodata & \nodata & \nodata \\
DO Peg & 22 07 30.6309 & +06 10 16.55 & 10.60 & 2.90 & V & B & 0 & 0.90 & \nodata & B8+[G4IV] & SA \\
EL Lac & 22 08 53.7317 & +42 16 20.967 & 12.40 & 3.60 & p & M & 0 & 0.64 & 12.53 & \nodata & \nodata \\
TV Cep & 22 09 53.9531 & +63 07 16.409 & 12.20 & 2.50 & p & M & 0 & 0.86 & 6.47 & K5V & \nodata \\
ER Lac & 22 19 03.1800 & +51 41 08.900 & 14.00 & 2.50 & p & M & 0 & \nodata & \nodata & \nodata & \nodata \\
BS Lac & 22 19 37.2753 & +44 17 03.117 & 13.50 & 2.70 & p & M & 0 & 0.94 & 6.69 & \nodata & \nodata \\
BR Cep & 22 27 17.1259 & +66 10 00.300 & 12.50 & 2.50 & p & M & 0 & 0.85 & 20.51 & A3 & \nodata \\
DY Lac & 22 47 19.2341 & +53 59 06.304 & 14.60 & 3.00 & p & M & 0 & 0.30 & 16.25 & \nodata & \nodata \\
EH Lac & 22 51 55.6894 & +51 25 14.120 & 13.60 & 3.20 & p & M & 0 & 0.34 & 14.92 & \nodata & \nodata \\
HI Lac & 22 56 48.6620 & +53 47 47.610 & 15.00 & 3.00 & p & M & 0 & 0.74 & 2.04 & \nodata & \nodata \\
BO And & 22 58 38.0000 & +45 31 52.000 & 13.40 & 2.90 & P & B & 0 & 0.34 & 37.6 & B8+[G6IV] & SA \\
V0341 Cas & 22 59 34.7018 & +56 23 09.759 & 14.70 & 2.60 & b & M & 0 & 0.52 & 8.77 & \nodata & \nodata \\
CU And & 23 01 01.5967 & +49 58 25.318 & 12.50 & 3.50 & p & M & 0 & 0.92 & 6.71 & \nodata & \nodata \\
V0570 Cas & 23 16 27.5700 & +59 48 18.194 & 13.80 & 2.70 & B & M & 0 & 1.27 & 0.67 & F-G & \nodata \\
X Gru & 23 19 42.3634 & -55 36 41.485 & 10.64 & 3.64 & V & B & 1 & 1.47 & \nodata & A0 & SA \\
Y Psc & 23 34 25.3848 & +07 55 28.524 & 10.10 & 3.00 & p & M & 1 & 2.30 & 26.97 & A3+K0IV & SA \\
V0442 Cas & 23 40 14.7999 & +53 57 33.990 & 13.20 & 3.80 & b & M & 0 & 0.56 & 21.85 & \nodata & \nodata \\
VZ Scl & 23 50 09.2550 & -26 22 52.701 & 15.60 & 2.80 & V & M & 0 & 1.78 & \nodata & pec(e) & S2C \\
QS Cas & 23 51 58.6213 & +56 02 31.891 & 13.60 & 2.90 & p & M & 0 & 0.34 & 17.32 & \nodata & \nodata \\
\enddata
\tablecomments{Eclipsing binary systems with primary minimum depth greater than or equal to 2.5. (1) General Catalog of Variable Stars designation, (2) Right ascension, (3) Declination, (4) Magnitude at maximum brightness, (5) Depth of primary minimum, (6) Photometric filter used to obtain the light curve, (7) Reference of target, maximum brightness, primary minimum, and filter (B is \cite{2004AA...417..263B} and M is \cite{2006AA...446..785M}), (8) Discrepancy flag, (9) Parallax, (10) Maximum bolometric luminosity, (11) Spectral type, (12) Evolutionary class. Columns (2) and (3) are extracted from SIMBAD and refer to epoch J2000, column (11) is from \cite{2004AA...417..263B}, \cite{2006AA...446..785M}, and \cite{2013AN....334..860A}, and column (9) and (10) are from the \textit{Gaia} DR2 database. If, for a given target, the catalogs report significantly different values of Dmag in the same photometric filter, we exclude that target from the analysis, but we report them here for completeness. A value of 0 and 1 for the discrepancy flag represents consistent and discrepant Dmag values respectively.}
\end{deluxetable}
\end{longrotatetable}

\end{document}